\documentclass[a4paper,11pt]{article}
\usepackage[dvipdfmx]{graphicx}
\usepackage{ulem}
\usepackage{jheppub} 

\usepackage[T1]{fontenc} 

\usepackage{slashed}
\usepackage{subfigure}

\def\nn{\nonumber}
\def\gev{{\rm GeV}}
\def\tev{{\rm TeV}}
\def\mev{{\rm MeV}}
\def\kev{{\rm keV}}
\def\ev{{\rm eV}}

\newcommand{\lsim}{
\mathrel{\hbox{\rlap{\hbox{\lower4pt\hbox{$\sim$}}}\hbox{$<$}}}}
\newcommand{\gsim}{
\mathrel{\hbox{\rlap{\hbox{\lower4pt\hbox{$\sim$}}}\hbox{$>$}}}}

\title{\boldmath Right-handed neutrino dark matter under the $B-L$ gauge interaction}

\author[a]{Kunio Kaneta}
\author[b]{Zhaofeng Kang}
\author[a,1]{Hye-Sung Lee,\note{Corresponding author.}}
\affiliation[a]{Center for Theoretical Physics of the Universe, Institute for Basic Science, Daejeon 34051, Korea}
\affiliation[b]{School of Physics, Korea Institute for Advanced Study, Seoul 02455, Korea}
\emailAdd{kaneta@ibs.re.kr}
\emailAdd{zhaofengkang@gmail.com}
\emailAdd{hlee@ibs.re.kr}

\abstract{We study the right-handed neutrino (RHN) dark matter candidate in the minimal $U(1)_{B-L}$ gauge extension of the standard model.
The $U(1)_{B-L}$ gauge symmetry offers three RHNs which can address the origin of the neutrino mass, the relic dark matter, and the matter-antimatter asymmetry of the universe.
The lightest among the three is taken as the dark matter candidate, which is under the $B-L$ gauge interaction.
We investigate various scenarios for this dark matter candidate with the correct relic density by means of the freeze-out or freeze-in mechanism.
A viable RHN dark matter mass lies in a wide range including keV to TeV scale.
We emphasize the sub-electroweak scale light $B-L$ gauge boson case, and identify the parameter region motivated from the dark matter physics, which can be tested with the planned experiments including the CERN SHiP experiment.}

\preprint{CTPU-16-17}

\begin{document} 
\maketitle
\flushbottom

\section{Introduction}
\label{sec:introduction}
In understanding nature, the gauge symmetry and its spontaneous breaking play a core role.
The standard model (SM) of particle physics is an extremely successful model so far in explaining the data.
Needless to say, its beauty is ascribed to the gauge principle which not only regulates the interactions among particles but also organizes the content of particles by means of the anomaly cancellation conditions. 

There are, however, various issues that the SM fails to address.
For instance, although the existence of the dark matter (DM) is quite certain to explain many independent astrophysical observations, it is convinced that the dark matter does not belong to the SM, and its identity has been still unknown.
Due to the fact that the neutrinos are massive, there plausibly exist their chiral partners, the right-handed neutrinos (RHNs), which are also not a part of the SM.
Unlike the other SM fermions, they can be Majorana particles which can exploit the seesaw mechanism to explain their small masses \cite{seesaw}. 
Through the seesaw mechanism, the RHN can stay effectively as a sterile neutrino, decoupled from the active neutrinos.

The RHN, which is neutral under the SM gauge symmetries, has a potential to be a viable dark matter candidate. 
This has been realized in the $\nu$ minimal standard model ($\nu$MSM) \cite{Asaka:2005an,Asaka:2005pn},\footnote{For some reviews of the $\nu$MSM and the light sterile neutrino dark matter physics, see Refs.~\cite{Boyarsky:2009ix,Alekhin:2015byh,Adhikari:2016bei}.} which sets the lightest RHN ($N_1$) mass around keV scale such that it can be naturally long-lived against its decay, $N_1\to\nu\gamma$, induced through the mixing between $N_1$ and active neutrinos, where the mixing angle is conventionally denoted by $\theta_1$. 
The framework of the $\nu$MSM can also address the baryon asymmetry of the universe (BAU) \cite{Asaka:2005pn} through the GeV scale RHNs and active neutrino oscillations \cite{Akhmedov:1998qx}.

In the $\nu$MSM, the sterile neutrino DM can be produced through the mixing between the $N_1$ and the active neutrinos, which is known as Dodelson-Widrow mechanism \cite{Dodelson:1993je} (see also Refs.~\cite{Barbieri:1989ti,Asaka:2006nq}). 
However, the non-observation of the $X$-ray signal from the $N_1$ decay ($N_1 \to \nu\gamma$) \cite{Essig:2013goa} and the phase space density constraint on the $N_1$ mass \cite{Horiuchi:2013noa} excluded this simple approach (for the Lyman-$\alpha$ forest constraint, see, e.g., Ref.~\cite{Viel:2013apy}) except for turning to the resonant effect which requires an anomalously large lepton asymmetry~\cite{Shi:1998km}. 
As another way out, introducing extra interactions can provide a viable dark matter production mechanism that is independent of the mixing angle $\theta_1$. 
For instance, the $N_1$ can be produced by the decay of a scalar particle \cite{Asaka:2006ek,Merle:2013wta} through the freeze-in mechanism \cite{FIMP}.
(For a discussion on the freeze-in scenario for the hidden sector dark matter that communicates with our sector through the kinetic mixing and/or the scalar mixing, see Ref.~\cite{Chu:2011be}.)

In this paper, we present the minimal $U(1)_{B-L}$ gauge extension of the SM with the RHN dark matter candidate, which we call the $U(1)_{B-L}$ extended $\nu$MSM or the U$\nu$MSM.
We also explore a comprehensive picture of the sterile neutrino DM candidate in this model.
In the light of the success of the gauge principle in the SM, the $U(1)_{B-L}$ gauge symmetry is expected to play a similar role for the DM\footnote{The $U(1)_{B-L}$ gauge symmetry is also attractive for asymmetric dark matter scenarios (for instance, see Ref.~\cite{ADM}).}.
In fact, due to the anomaly cancellation conditions, the $U(1)_{B-L}$ regulates the number of the RHNs to be three.
The lightest RHN can be a DM candidate with its mass scale from keV to TeV, or even higher.
The other two RHNs may be responsible for the BAU, which will be studied elsewhere.
The interaction can be mediated by both a $B-L$ gauge boson $Z'$ and an associated scalar $S$ that is associated with the spontaneous symmetry breaking.
To gain the control in the number of free parameters, we will consider only the $Z'$ interaction in this work, unless specifically stated, which is valid in the limit the $S$ is heavy enough and/or inefficiently communicate with the SM sector so that its contribution to the DM production is greatly suppressed.
The new gauge interaction can play an important role in the sterile neutrino production especially via the $Z'$ mediated freeze-in mechanism, and provide distinguishable implications that can be tested experimentally.

There are some related works such as Refs.~\cite{Okada:2010wd,Okada:2012sg,Okada:2016gsh}.
They impose a $Z_2$ protective symmetry on some sterile neutrino while requiring two others to accommodate realistic neutrino phenomenology.
In this scenario, the sterile neutrino can be an ordinary cold DM candidate around the weak scale, i.e., it has a weak interaction (say, the $U(1)_{B-L}$ gauge interaction) and gains a correct relic density via the conventional freeze-out mechanism.
In our study, the most interesting case (also the main case) actually is a very light RHN which does not necessarily call for a $Z_2$ protective symmetry, although for the sake of a global picture we also  include the heavy RHN dark matter case, which then may require a flavor symmetry as in Refs.~\cite{Okada:2010wd,Okada:2012sg,Okada:2016gsh}.
We also exploit the freeze-in mechanism to account for correct relic density for the RHN DM.
A scalar DM candidate in a similar framework was studied in Ref.~\cite{Guo:2015lxa}.
We also note a larger gauge group $SU(3)_C\times SU(2)_L\times SU(2)_R\times U(1)_{B-L}$ based on the Left-Right gauge symmetry was considered before \cite{Bezrukov:2009th,Nemevsek:2012cd}.
Heavy gauge bosons ($W^\pm_R$ and $Z'$) and usual freeze-out production method with a dilution was used, which is a different approach from ours.

The rest of this paper is organized as follows.
In Sec.~\ref{sec:model}, we describe our framework, the U$\nu$MSM.
In Sec.~\ref{sec:darkmatter}, we discuss possible DM production scenarios and the relevant constraints on the model.
In Sec.~\ref{sec:implications}, we discuss implications for various phenomena including the SHiP experiment.
In Sec.~\ref{sec:summary}, we summarize our study.

\section{\boldmath The framework of the U$\nu$MSM}
\label{sec:model}
Following the success of the gauge principle in the SM, we consider a model with the $U(1)_{B-L}$ gauge symmetry as a minimal choice in terms of the matter contents, which offers three RHNs $N_i$, a $U(1)_{B-L}$ gauge boson $Z'$, and a single scalar $\Phi_S$ being responsible for spontaneous breakdown of $U(1)_{B-L}$. 
The Lagrangian of the U$\nu$MSM is given by
\begin{eqnarray}
  {\cal L} 
  =&& {\cal L}_{\rm SM} + i\overline N_i\slashed{D}N_i-\left(y_{\alpha i}\overline L_\alpha N_i \tilde \Phi_H + \frac{\kappa_i}{2} \Phi_S \overline{N^C_i} N_i + h.c.\right)\nn\\
  &&+ |D_\mu \Phi_S|^2 - V(\Phi_H, \Phi_S) - \frac{1}{4}Z'_{\mu\nu}Z'^{\mu\nu} + \frac{\epsilon}{2} Z'_{\mu\nu}B^{\mu\nu},
\end{eqnarray}
where $\alpha=e,\mu,\tau$, $i=1,2,3$, and $D_\mu=\partial_\mu-ig_{B-L} Q' Z'_{\mu}$ with $g_{B-L}$ and $Q'$ being the $B-L$ gauge coupling and $B-L$ charge ($Q'=-1$ for the SM leptons and $N$'s, $Q'=1/3$ for the SM quarks, $Q'=2$ for $\Phi_S$).
$Z'_{\mu\nu}$ is the field strength of the $Z'$.
We take four-component fermion notations, by which $N_i$ represents a four-component fermion having only the right-handed part.

The gauge kinetic mixing of $(\epsilon / 2) Z'_{\mu\nu}B^{\mu\nu}$ is highly constrained, and for the simplicity we take it zero in this paper.
The gauge kinetic mixing \cite{Holdom:1985ag} has been a great source of research interests in the past decade \cite{Essig:2013lka} and also branched out some variant forms such as the one in Ref.~\cite{Davoudiasl:2012ag}.
See Ref.~\cite{Lee:2016ief} for the details of the physics related to this term in the gauged $B-L$ model.

The Higgs potential is given by
\begin{eqnarray}
  V(\Phi_H,\Phi_S) 
  =&&\frac{\lambda_H}{2}(|\Phi_H|^2-v_H^2)^2+\frac{\lambda_S}{2}(|\Phi_S|^2-v_S^2)^2\nn\\
  &&+\lambda_{HS}(|\Phi_H|^2-v_H^2)(|\Phi_S|^2-v_S^2),
\end{eqnarray}
where $\Phi_H$ and $\Phi_S$ develop the vacuum expectation values (VEVs), $\langle \Phi_H\rangle=v_H$ and $\langle \Phi_S\rangle=v_S$, so that the electroweak and the $B-L$ gauge symmetries are spontaneously broken.
After diagonalizing the mass matrix, we obtain the masses
\begin{eqnarray}
  M_H^2 &\simeq& 2\lambda_H v_H^2 - 2\lambda_{HS}v_Hv_S\theta,\\
  M_S^2 &\simeq& 2\lambda_S v_S^2 + 2\lambda_{HS}v_Hv_S\theta,\label{eq:Ms}
\end{eqnarray}
for the physical states $H$ and $S$, respectively, where the mixing angle is given by $\tan2\theta=2\lambda_{HS}v_Hv_S/(\lambda_Hv_H^2-\lambda_Sv_S^2)$.
The VEV of $\Phi_S$ gives the mass of $Z'$ and $N_i$ as follows:
\begin{eqnarray}
  M_{Z'}^2 &=& 8g_{B-L}^2v_S^2,\label{eq:MZp}\\
  M_{N_i} &=& \kappa_i v_S.\label{eq:MN}
\end{eqnarray}
The coupling $\kappa_i$ is in general a complex value.
Our following discussion is, however, independent from its CP phases, and thus we take $\kappa_i$ as a real parameter in what follows.
The $N_2$ and $N_3$ are not directly related to the DM production and their masses are not bounded by the DM relic density as in the $\nu$MSM, as the resonant production through the large lepton asymmetry is not necessary in this model.

The dominant decay mode is $N_1 \to 3 \nu$ given by \cite{Pal:1981rm,Barger:1995ty}
\begin{eqnarray}
\Gamma_{N_1 \to 3 \nu} = \frac{G_F^2 M_{N_1}^5}{96 \pi^3} \sin^2\theta_1 .
\end{eqnarray}
Requiring the $N_1$ lifetime is longer than the universe age ($\tau_U \sim 13.7 \times 10^9$ years), we get the following constraint.
\begin{eqnarray}
\Big(\frac{M_{N_1}}{\kev}\Big)^3 \Big( \frac{\sum_{\alpha} |y_{\alpha 1}|^2}{5.5\times10^{-16}} \Big) \lsim 1
\end{eqnarray}
where we have used $\theta_1^2 \simeq \sum_{\alpha} |y_{\alpha 1}|^2 v_H^2  / M_{N_1}^2$ from the the see-saw mechanism.
Thus, the low mass of the $N_1$ (not too larger than the $\ev$ scale) can satisfy the DM lifetime constraint easily, but the heavier $N_1$ would require $\sum_{\alpha} |y_{\alpha 1}|^2 \ll 1$ to be sufficiently stable.

Although the heavier the $N_1$ DM may mean the less natural setup, we will include the heavier $N_1$ in our study that expands the relevant phenomenology significantly (e.g., see Sec.~\ref{sec:implications}).
As a matter of fact, the $N_1$ will be stable as long as the $\sum_{\alpha} |y_{\alpha 1}|^2 \simeq 0$.
In this limit, which might invoke a flavor symmetry like Refs.~\cite{Okada:2010wd,Okada:2012sg,Okada:2016gsh}, the lightest neutrino would be massless ($m_{\nu_1} = 0$) or almost massless, which is still consistent with the experimental constraints \cite{Agashe:2014kda}.
Throughout the rest of this paper, we will discuss in the zero $N_1$ mixing angle ($\theta_1 = 0$) limit, which also allows us to leave out of account the constraints from the $X$-ray observations with the $N_1 \to \gamma + \nu$ process.

\section{Dark matter production and constraints}
\label{sec:darkmatter}
We now turn to discussing how the $B-L$ gauge boson $Z'$ makes an impact on the $N_1$ dark matter production. 
The dark matter scenario drastically changes, depending on whether the $Z'$ can decay into the dark matters ($M_{Z'}>2M_{N_1}$) or not ($M_{Z'}<2M_{N_1}$).

In the rest of this section, we will approach the dark matter issues from very general points.
First, we will discuss how and where the $N_1$ and $Z'$ can be thermalized (Sec.~\ref{subsec:thermalization}).
Then, we will discuss various constraints including the Big Bang nucleosynthesis (BBN), lab experiments, and astrophysical bounds (Sec.~\ref{subsec:constraints}) before we discuss the dark matter relic density.
Although some of the discussions and constraints may not be directly relevant to the parameter region that gives the right relic density for the $N_1$, it might be still instructive to have them as they might be relevant when we consider somewhat altered scenario such as the late time entropy injection.
In Sec.~\ref{subsec:dm0}, we briefly go over the issues for the keV scale $N_1$ dark matter for the thermal production.
We discuss mainly the non-thermal $N_1$ dark matter production for the $M_{Z'} > 2 M_{N_1}$ ($M_{Z'} < 2 M_{N_1}$) case in Sec.~\ref{subsec:dm1} (Sec.~\ref{subsec:dm2}).

\subsection{Thermalization of the $N_1$ and $Z'$}
\label{subsec:thermalization}
Before heading towards the production of the correct relic density of the $N_1$, we describe how the dark sector, the dark matter as well as its portal $Z'$, is thermalized.
For the thermalization of the $N_1$ and $Z'$, the relevant reactions among the $N_1$, $Z'$ and the SM particles are (a) $N_1\overline N_1\leftrightarrow f\overline f$, (b) $Z'Z'\leftrightarrow f\overline f$, and (c) $N_1\overline N_1\leftrightarrow Z'Z'$, of which the reaction rates are denoted by $r_a$, $r_b$, and $r_c$, respectively. The relevant formulae are given in Appendix \ref{sec:appendix}. 
If $r_i$ ($i=a,b,c$) is larger than the Hubble expansion parameter, $H=(g_*\pi^2/90)^{1/2}(T^2/M_{\rm Pl})$ with $M_{\rm Pl}\simeq 2.4\times10^{18}~\gev$ being the reduced Planck mass, at some time, the $N_1$ and/or $Z'$ enter the thermal bath (reaching the relative chemical equilibrium of the SM sector and/or dark sector).
In the following discussion, we take the numbers of degrees of freedom for the energy density and the entropy density to be the same value $g_*$ since they are very close, and $g_*$ is evaluated as a function of the temperature according to Ref.~\cite{Hindmarsh:2005ix}.

It should be noted that $N_1\overline N_1\leftrightarrow Z'Z'$ mediated by s-channel $S$ also exists.
As we will discuss later, however, $S$ can be always heavier than the $N_1$ and $Z'$ in the parameter regions of our interest, and this process will be suppressed as we will take a very heavy $S$.
For other possible processes, $N_2\overline N_2,~N_3\overline N_3\leftrightarrow N_1\overline N_1$ mediated by $S$ may become significant when $\kappa_i$ is strong. 
In such a case, $SS\leftrightarrow N_1\overline N_1$ may also be relevant for the thermalization. 
On the other hand, as we will see, we can take $M_{N_1},M_{Z'}<M_{N_2},M_{N_3},M_S$ in the parameter region of our interest, and these processes can be omitted by taking a specific reheating temperature $T_R$ as ${\rm max}\{M_{N_1},M_{Z'}\}\lesssim T_R\lesssim {\rm min}\{M_{N_2},M_{N_3},M_S\}$. 
In what we follow we take this case for the sake of simplicity.
In order to focus on the $Z'$ interaction, we turn off the other possible reactions involving scalars, such as $HH,SS,SH\leftrightarrow N_1\overline N_1$, by taking $S$ very heavy and $\lambda_{HS}$ vanishingly small in a similar way to Ref. \cite{Asaka:2006ek}.

\begin{figure*}[t]
\begin{center}
\subfigure[]{\label{fig:10keV_1}
\includegraphics[width=0.45\textwidth,clip]{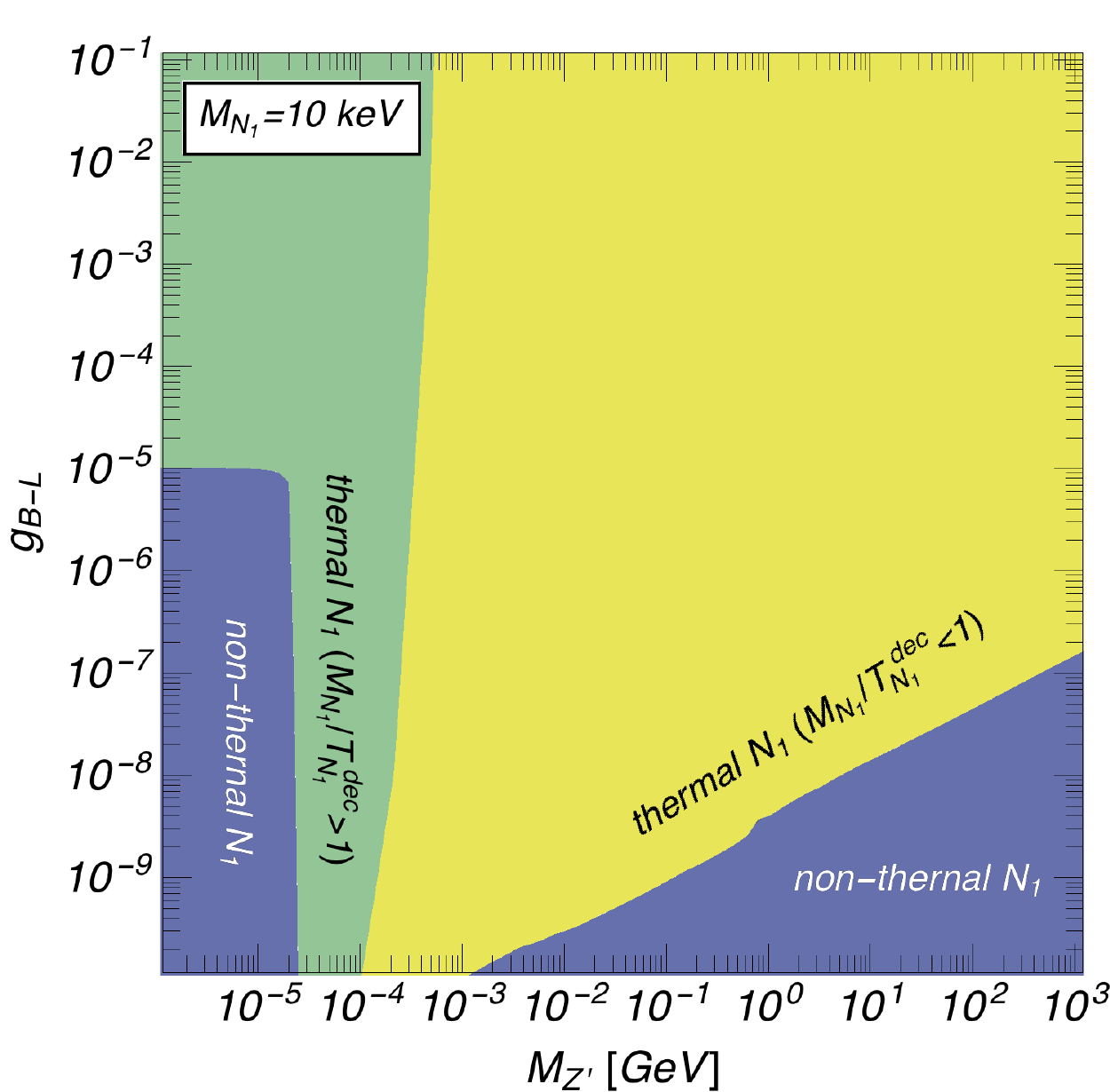}} ~~
\subfigure[]{\label{fig:10keV_2}
\includegraphics[width=0.45\textwidth,clip]{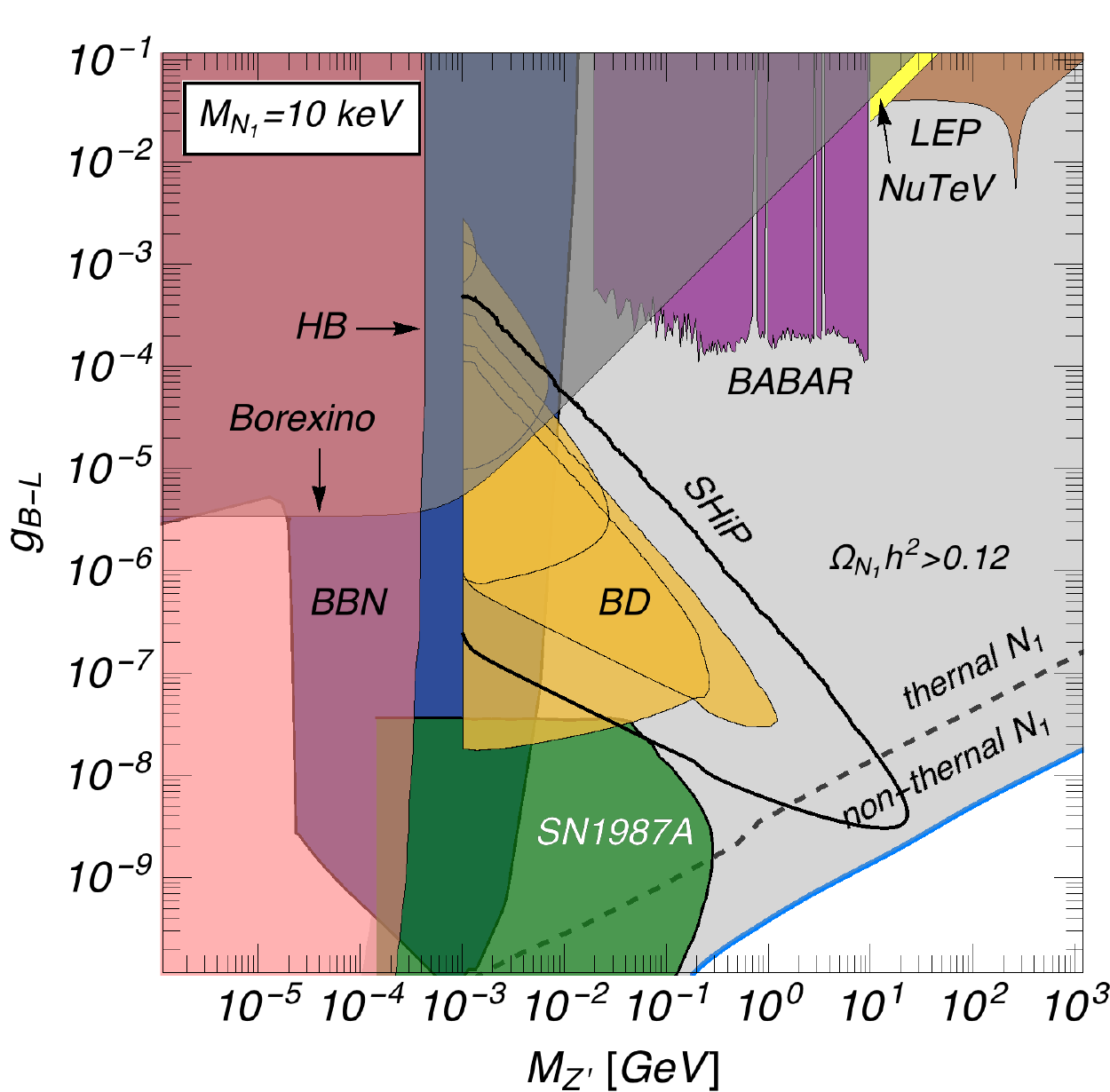}}
\end{center}
\caption{The production of the $N_1$ of $10~\kev$ depending on the $B-L$ gauge boson mass and coupling (a) without and (b) with various experimental constraints (including the DM relic density) imposed.
In the (a), the non-relativistic (light green), the relativistic (light yellow), and the non-thermal (deep blue) regions are indicated. The blue curves indicate the points where the DM relic density $\Omega_{N_1} h^2 = 0.12$ is satisfied.}
\label{fig:10keV}
\end{figure*}

Figure~\ref{fig:10keV_1} shows whether the $N_1$ is thermalized or not depending on the $Z'$ mass and coupling, where we take $M_{N_1}=10$ keV for an illustration purpose. 
In the deep blue regions, the $N_1$ never enters the thermal bath; in the other regions, the $N_1$ becomes thermal at some time. 
For the thermal $N_1$, there are two distinct regions depending on if the $N_1$ is relativistic (hot or warm dark mater case) or non-relativistic (cold dark matter case) at its decoupling temperature $T_{N_1}^{\rm dec}$ which is evaluated by $r_{a,b}(T_{N_1}^{\rm dec})= H(T_{N_1}^{\rm dec})$.
In the light yellow region, the $N_1$ satisfies $M_{N_1}/T_{N_1}^{\rm dec}<1$, namely, it is a relativistic particle, while in the light green region, the $N_1$ is a non-relativistic particle.

In thermalization of the $N_1$, the reaction rate $r_a$ is the dominant contribution to take the $N_1$ into the thermal equilibrium with the SM particles.\footnote{If $Z'$ is thermalized via the reaction (b), the $N_1$ can be also thermalized via the reaction (c). 
This contribution is, however, subdominant for $M_{Z'}>2M_{N_1}$ as the reaction (a) with the on-resonance enhancement dominates.}
As mentioned in the beginning of this section, the DM production is sensitive to the critical line $M_{Z'}\sim 2M_{N_1}$.
When $M_{Z'}>2M_{N_1}$, $r_a$ is enhanced by the on-resonance contribution, and thus the $N_1$ is thermalized even when $g_{B-L}$ is very small; there is also a region in the bottom right corner of the parameter space where the $N_1$ is not thermalized because the mediator $Z'$ is too heavy and suppresses the reaction rate.
On the other hand, $r_a$ gets suppressed for $M_{Z'}<2M_{N_1}$ since the process (a) becomes off-resonance, and the required $g_{B-L}$ for thermalization becomes larger. 
Figure~\ref{fig:10keV_1} would not change even if there is a late time entropy injection, and clearly illustrates the distinction between $M_{Z'} < 2 M_{N_1}$ region and $M_{Z'} > 2 M_{N_1}$ region.

\subsection{Collecting relevant constraints}
\label{subsec:constraints}
In Fig.~\ref{fig:10keV_2}, we collect relevant constraints in the $g_{B-L}$ and $M_{Z'}$ parameter space, for a choice of $M_{N_1} = 10$ keV (which is considered to be a conservative value for the lowest $M_{N_1}$ \cite{Horiuchi:2013noa}).
The constraint from Big Bang nucleosynthesis (BBN) can eliminate a large part of the parameter space shown in dark blue.
The existence of additional relativistic degrees of freedom can speed up the expansion of the universe, which leads to the earlier decoupling of the active neutrinos, and hence a higher yield of $^4$He and so on.
The extra radiation density is included in the conventional parametrization of
\begin{eqnarray}
  \rho &=& N_{\rm eff}\frac{7}{8}\left(\frac{4}{11}\right)^{4/3}\rho_\gamma ,
\end{eqnarray}
where $\rho_\gamma$ is the photon energy density, and $N_{\rm eff}$ counts 3 for three active neutrinos.\footnote{Here we ignore the flavor dependence of the neutrino decoupling temperature, and take $T_\nu^{\rm dec}\sim 1~\mev$. In reality, $\nu_\mu$ and $\nu_\tau$ might decouple before $\nu_e$, which would induce a small correction to $N_{\rm eff}$.}

In our case, the deviation from 3 contains the contributions from the $N_1$ and $Z'$ (if it is relativistic at $T_{\nu}^{\rm dec}\sim1~\mev$), which is given by
\begin{eqnarray}
  \Delta N_{\rm eff}
  &\simeq&
  \frac{12}{7}\left[\frac{g_*(T_{\nu}^{\rm dec})}{g_*(T_{Z'}^{\rm dec})}\right]^{4/3}+\left[\frac{g_*(T_{\nu}^{\rm dec})}{g_*(T_{N_1}^{\rm dec})}\right]^{4/3},
\end{eqnarray}
where $g_*(T_{\nu}^{\rm dec})=10.75$.
By demanding $\Delta N_{\rm eff}<1$ \cite{Ade:2015xua}, we obtain the exclusion region shaded in dark blue for the range of $1~\mev \lesssim M_{Z'}\lesssim 10~\mev$ in Fig.~\ref{fig:10keV_2}.
For masses $2 M_{N_1}\lesssim M_{Z'}\lesssim 1~\mev$, we impose that the $N_1$ enters the thermal bath after $T\sim 1~\mev$ so that the $N_1$ does not affect the SM neutrino decoupling \cite{Ahlgren:2013wba,Heeck:2014zfa}, which leads to the bound for the coupling, $g_{B-L}\gtrsim 3\times 10^{-9}-10^{-10}$.\footnote{When the thermalization temperature of the $N_1$ is lower than the temperature at which the BBN is completed, observations of the light elements can not give any constraints. On the other hand, when the thermalization of the $N_1$ occurs after the recombination ($T \sim 0.1$ eV), the thermalized $N_1$ may leave an imprint on the cosmic microwave background. This temperature range is beyond the scope of this paper though.}
For $M_{Z'} \lesssim 2 M_{N_1}$ and $g_{B-L} < \text{several} \times 10^{-6}$, only the thermal $Z'$ contributes to $\Delta N_{\rm eff}$ because the $N_1$ is non-thermal.

The other individual constraints shown in Fig.~\ref{fig:10keV_2} are following.\footnote{We did not take into account the $Z' \to N_1 \overline N_1$ branching ratio for the BABAR, BD, SHiP, LEP bounds, which depend on it, and these bounds are taken as the same as Figs.~\ref{fig:fig_Oh2} - \ref{fig:fig_focus}. The change will be small nevertheless.}
\begin{enumerate}
\item {\it LEP experiments.}
  The high mass regions are sensitive to the LEP experiments which give the exclusion limit depicted by the brown region \cite{Jeong:2015bbi}.
  The constraint for the contact interactions \cite{Carena:2004xs} is valid only for the $M_{Z'}$ much larger than the collision energy at LEP, 209 GeV, while the initial state photon radiation process, $e^+e^-\to\gamma\nu\bar\nu$ \cite{Achard:2003tx}, can be used for the $M_{Z'}$ lower than the collision energy.

\item {\it BABAR experiments.}
For $20~\mev < M_{Z'} < 10~\gev$, the BABAR experiments give the stringent bound from $e^+e^-\to\gamma Z'$ followed by $Z'\to e^+e^-/\mu^+\mu^-$ at around $\Upsilon$ resonances \cite{Lees:2014xha}, which is represented by the purple region.

\item {\it Beam dump (BD) experiments.}
The orange regions are excluded by the electron and proton BD experiments, where the regions from top to bottom correspond to E774 \cite{Bross:1989mp}, E141 \cite{Riordan:1987aw}, Orsay \cite{Davier:1989wz}, $\nu$-Cal I (proton bremsstrahlung) \cite{Blumlein:2013cua}, E137 \cite{Bjorken:1988as}, respectively.
The black solid curve shows the expected reach of the SHiP experiment based only on the proton bremsstrahlung \cite{Gorbunov:2014wqa,Alekhin:2015byh}, which we will discuss in Sec.~\ref{sec:implications}.
We have followed the method in Ref.~\cite{Andreas:2012mt} to calculate the bounds from the electron beam dump experiments.
For the proton beam dump experiments, the relevant calculation is shown in Refs.~\cite{Gorbunov:2014wqa,Blumlein:2013cua}.

\item {\it $\nu-e$ scattering at Borexino.}
The Borexino experiment has reported the interaction rate of neutrino-electron scattering from 867 keV $^7$Be solar neutrinos \cite{Bellini:2011rx}.
The observed value is consistent with the SM prediction, which gives the bound denoted by the dark gray region, by imposing that the ratio between the cross section involving $Z'$ and the SM contributions should not exceed the maximum error \cite{Harnik:2012ni}. 
This constraint is very powerful as it applies to a wide region of $M_{Z'}$.
See also Ref.~\cite{Bilmis:2015lja} for the similar level of the constraint from the $\bar\nu-e$ scattering based on the reactor experiments.
  
\item{\it $\nu-q$ scattering at NuTeV.}
The mass range of $Z'$ above 10 GeV is constrained by the neutrino-nucleon scatterings.
The NuTeV experiment measured $\nu_\mu(\bar \nu_\mu)-q$ scattering, where $\nu_\mu$ and $\bar\nu_\mu$ were provided by the beamline at the Fermilab \cite{Zeller:2001hh}.
Since there is relatively large systematic errors, we take a conservative limit: $M_{Z'}/g_{B-L}>0.4$ TeV \cite{Escrihuela:2011cf,Heeck:2014zfa}, which is depicted by the light yellow region.

\item {\it Horizontal-branch (HB) stars.}
For the lighter $Z'$, the energy loss rate of the stars in the globular clusters gives the more restrictive constraints, where the larger energy loss shortens the lifetime of the stars, and hence the observed population of the stars would be changed \cite{Raffelt:2000kp}.
Here, we show the constraint from HB stars represented by the red region \cite{Redondo:2013lna}.

\item {\it Supernova 1987A (SN1987A).}
The green region shows the constraint from the supernova explosion.
The extra light particle taking energies from the center of the supernova can affect the signal duration of the neutrinos \cite{Raffelt:2000kp}, in which the energy loss argument puts the bound \cite{Dent:2012mx}.
An updated constraint \cite{Kazanas:2014mca}, although not taken in our paper, is similar to the one in Ref.~\cite{Dent:2012mx} for the parameter regions we plot.
(Cf. For a discussion on the potential way out, called the Chameleon effect, see Ref.~\cite{Nelson:2007yq}.)
\end{enumerate}
The latest LHC bound on the $Z'$ through the Drell-Yan process comes into the region above TeV scale \cite{Okada:2016gsh}, which is beyond the region of our interest, and we omit it in the figure.

\subsection{Thermal production of the keV scale $N_1$}
\label{subsec:dm0}
As a warm-up, we first consider a well-known case that the $N_1$ is around the keV scale, specifically 10 keV, which can be a candidate for a warm dark matter.\footnote{A dedicated analysis on whether the $N_1$ is warm, by calculating its free stream, can be found in Ref.~\cite{Merle:2013wta}.} 
As one can see from Fig.~\ref{fig:10keV_1}, the $N_1$ can be thermalized in the bulk space of the $M_{Z'}-g_{B-L}$ plane, and we concentrate on this case.

The $N_1$ that once entered the plasma can be a warm or cold relic, depending on its mass and the decoupling temperature. 
The light yellow region in Fig.~\ref{fig:10keV_1} indicates that the $N_1$ is relativistic ($M_{N_1}/T_{N_1}^{\rm dec}<1$) at $T_{N_1}^{\rm dec}$, while it is non-relativistic ($M_{N_1}/T_{N_1}^{\rm dec}>1$) in the light green region, where $T_{N_1}^{\rm dec}$ is the decoupling temperature of the $N_1$.
When the $N_1$ is non-relativistic, $T_{N_1}^{\rm dec}$ is lower than $T_\nu^{\rm dec}$, and the BBN constraint excludes this region.
(The HB and SN1987A bounds also ruled out some part of this region independently.)

When the $N_1$ is relativistic at $T_{N_1}^{\rm dec}$, the relic abundance of the $N_1$ is given by
\begin{eqnarray}
  \Omega_{N_1} h^2 
  &=& 
  \frac{s_0 M_{N_1}}{\rho_ch^{-2}}\times \left.\frac{n_{N_1}}{s}\right|_{T_{N_1}^{\rm dec}}\nn\\
  &\simeq&
  110\times\left[\frac{M_{N_1}}{10~\kev}\right]\left[\frac{10.75}{g_*(T_{N_1}^{\rm dec})}\right],
  \label{eq:Oh2_rel}
\end{eqnarray}
where $n_{N_1}$ is the number density of the relativistic $N_1$, $n_{N_1}=(3/2)(\zeta(3)/\pi^2)T^3$, and $\rho_c=1.05368\times10^{-5} h^2 ~\gev~{\rm cm}^{-3}$ is the critical density of the universe. $s=(2\pi/45)g_*T^3$ and $s_0=2889.2~{\rm cm}^{-3}$ are the entropy density and its present day value. 
In this case, the abundance of the $N_1$ exceeds the observed value of the dark matter abundance $\Omega_{\rm DM}h^2\simeq 0.12$ \cite{Ade:2015xua}, and the universe is overclosed.
This parameter space is depicted by the gray region above the dashed curve in Fig.~\ref{fig:10keV_2}, excepting the non-relativistic region.

Such a large abundance could be diluted if we take into account the late time entropy production by, e.g, the decay of $N_{2,3}$ as studied in Refs.~\cite{Bezrukov:2009th,Nemevsek:2012cd} although they employed a different gauge extension.\footnote{Now the new singlet Higgs boson $S$ might be another candidate for late entropy production.
In order for this scenario to work, a careful analysis of the decay modes of the $S$ is necessary since the $S$ can decay into a pair of the $N_1$, which increases the $N_1$ number density.}
We note large parameter regions including a new window much below the weak scale can be viable in the case of the dilution, which has low energy laboratory implications.
This can be compared to the Refs.~\cite{Bezrukov:2009th,Nemevsek:2012cd} where only the weak scale or above was considered.
This is manifest in Fig.~\ref{fig:10keV_2}, which shows that BBN, BD and BABAR already excluded a large portion of the parameter space, and the SHiP experiment is able to cover more space.

\subsection{$M_{Z'}>2M_{N_1}$ case}
\label{subsec:dm1}
We here discuss the case of $M_{Z'}>2M_{N_1}$.
It is well known that when the $N_1$ is around the electroweak scale while the $Z'$ is at TeV scale, the $N_1$ can be a thermal relic dark matter.
This scenario was addressed in the context of the classically conformal models \cite{Iso:2009ss}, and collider signatures of such a heavy $Z'$ were studied in, e.g., Refs.~\cite{Okada:2016gsh,Basso:2008iv}.
We do not purse to study the thermal $N_1$ DM with a heavy $Z'$ in this paper.

On the other hand, there is another possibility that the $N_1$ is produced by the freeze-in mechanism \cite{FIMP}, where the $Z'$ is produced as an on-shell state, and subsequently decays into a pair of the $N_1$.
In this scenario, the $N_1$ never enters the thermal bath, and is produced by the annihilations of a pair of the SM particles and also a pair of the $Z'$ if it is thermalized.
This implies that the $Z'$ gauge coupling is very small compared to the thermal dark matter scenario.

We also require that the $N_1$ does not exist at the time when the universe is reheated up to the temperature $T_R$ after the inflation, namely $n_{N_1}(T_R)\simeq n_{N_1}(\infty)=0$, and thus the Boltzmann equation for $n_{N_1}$ is given by
\begin{eqnarray}
  \frac{dn_{N_1}}{dt}+ 3Hn_{N_1} = \frac{T}{64\pi^4}\int_{4M_{N_1}^2}^\infty ds~ \sigma v(s-4M_{N_1}^2)^{1/2} s K_1(\sqrt{s}/T) ,
  \label{eq:Boltzmann}
\end{eqnarray}
where $\sqrt{s}$ is the center of mass energy.
($K_1$ is the modified Bessel function of the first kind.)

For the annihilation cross section $\sigma v$, the process (a) is the dominant contribution, which is given by
\begin{eqnarray}
  \sigma v \simeq \frac{8}{3}g_{B-L}^4 \frac{s-4M_{N_1}^2}{M_{Z'}\Gamma_{Z'}}\delta(s-M_{Z'}^2) ,
  \label{eq:sigv-onshell}
\end{eqnarray}
where we have utilized the narrow width approximation.\footnote{We here consider the case that $T_R$ is sufficiently large compared to the masses of the $N_1$ and $Z'$. 
As another possibility, the scenario with $T_R<M_{Z'}$ was discussed in Ref.~\cite{Gelmini:2004ah}.}
Substituting Eq. (\ref{eq:sigv-onshell}) to the right hand side of Eq. (\ref{eq:Boltzmann}), we obtain
\begin{eqnarray}
  \frac{dY_{N_1}}{dT} = -\frac{45\sqrt{5}g_{B-L}^5}{8\sqrt{2}\pi^5}\frac{M_{\rm Pl}M_{Z'}^4}{g_*^{3/2}\Gamma_{Z'}T^5}K_1(M_{Z'}/T) \times\left[1-\frac{4M_{N_1}^2}{M_{Z'}^2}\right]^{3/2} ,
  \label{eq:Boltzmann2}
\end{eqnarray}
where we have used the yield $Y_{N_1}\equiv n_{N_1}/s$ and $d/dt = -HT~d/dT$, and take $g_*$ as a constant in the following.
By replacing $T$ with $x\equiv M_{N_1}/T$ and integrating $x$ from $0$ to $\infty$ in Eq.~\eqref{eq:Boltzmann2}, we end up with the non-thermal abundance
\begin{eqnarray}
  \Omega_{N_1}^{\rm nt}h^2
  &=&
  \frac{s_0 M_{N_1}Y_{N_1}^{\rm nt}}{\rho_ch^{-2}} \nn\\
  &\simeq&
  0.12\times\left[\frac{100}{g_*}\right]^{3/2}\left[\frac{g_{B-L}}{5.1\times10^{-12}}\right]^2\left[\frac{7}{C_f}\right]\left[\frac{f(\tau)}{0.19}\right],
  \nn\\
  \label{eq:Oh2nt_res}
\end{eqnarray}
where $f(\tau)=\tau(1-\tau^2)^{3/2}$ with $\tau=2M_{N_1}/M_{Z'}$ taking $0<\tau<1$, and $f(\tau)$ takes the maximal value $f(\tau)\simeq 0.19$ at $\tau=2/5$.
We also approximate the total decay width as $\Gamma_{Z'}\sim C_f g_{B-L}^2/(12\pi)M_{Z'}$ where $C_f$ is a coefficient in taking massless limit for final state particles.
If $Z'$ decays into all the SM fermions (and $N_1$), $C_f$ becomes 7.
We will approximate our results using $C_f = 7$, and the parameter region where the right DM relic density is satisfied will be slightly changed if we use the exact values.

In Fig.~\ref{fig:10keV_2}, we also depict the region of $\Omega_{N_1}^{\rm nt}h^2>0.12$ as the gray region below the dashed curve.
Therefore, the gauge coupling should be extremely small in order to obtain the observed dark matter abundance in this case, and it is quite challenging to test such a feebly interacting $Z'$ experimentally.

\begin{figure*}[t]
\begin{center}
\subfigure[]{\label{fig:fig_Oh2_1}
\includegraphics[width=0.45\textwidth,clip]{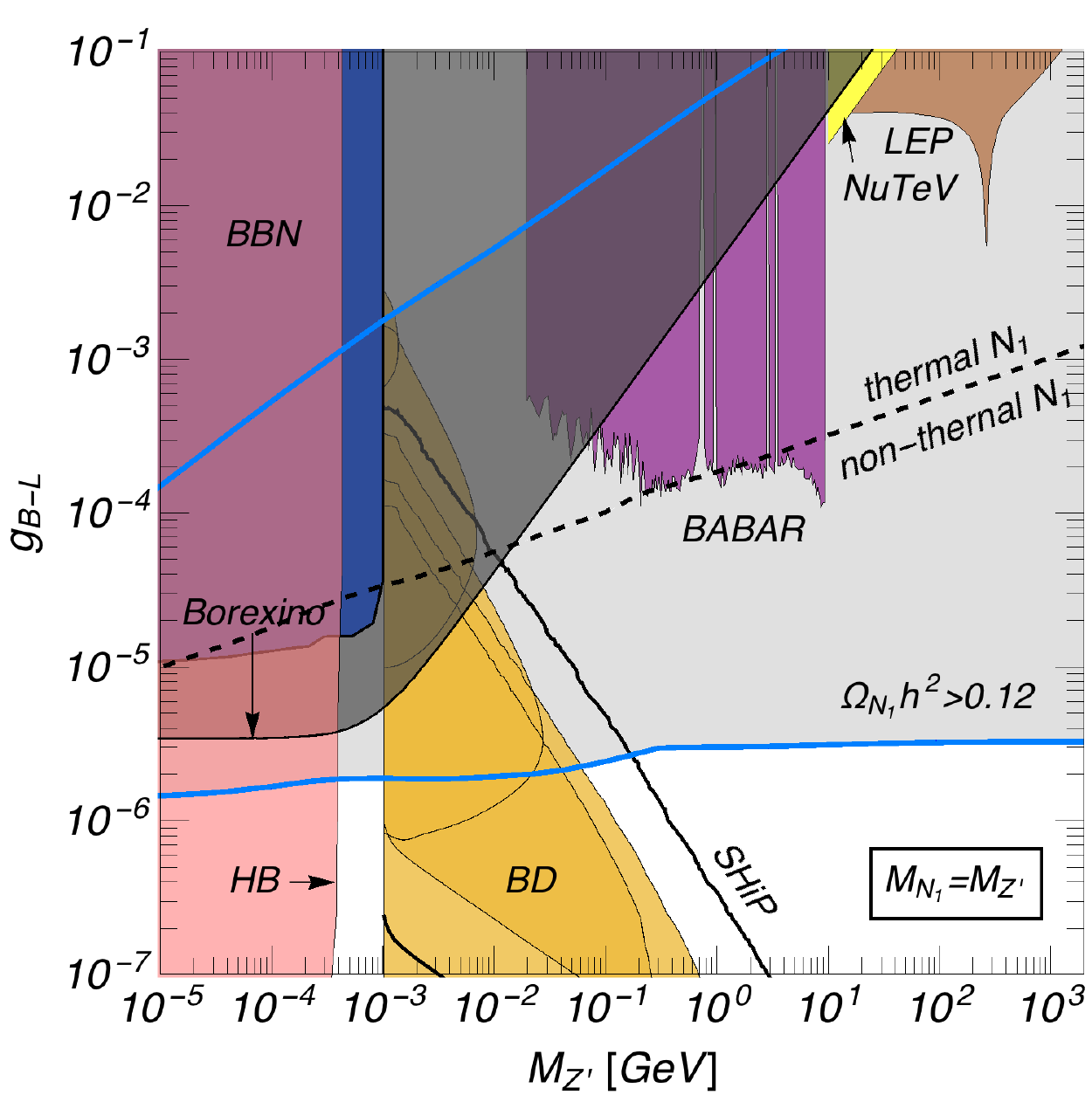}} ~~
\subfigure[]{\label{fig:fig_Oh2_2}
\includegraphics[width=0.45\textwidth,clip]{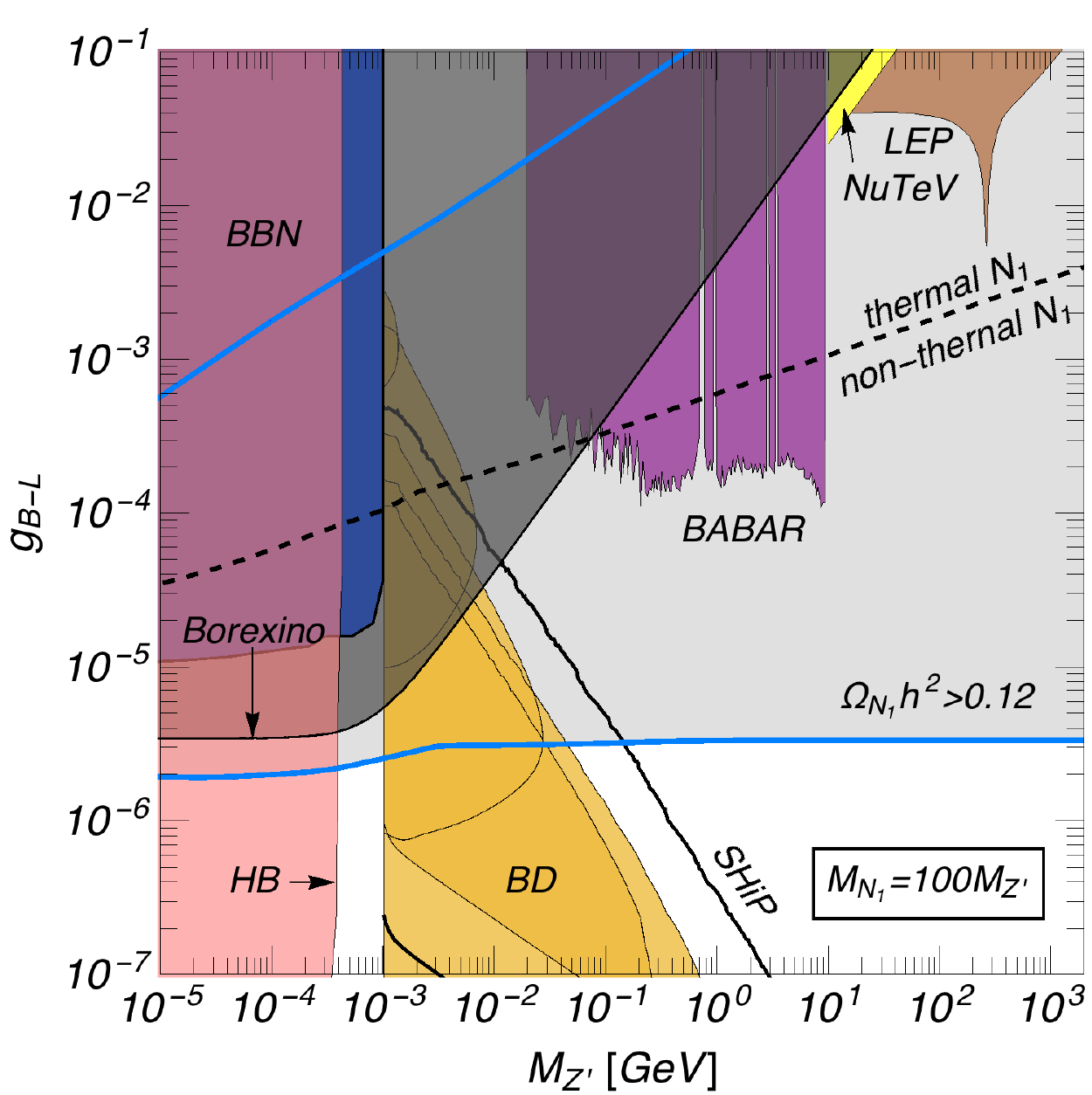}}
\end{center}
\caption{The $N_1$ dark matter abundance and various constraints on the gauge coupling and the mass of the $Z'$ for a couple of $M_{N_1} > M_{Z'} / 2$ cases: (a) $M_{N_1} = M_{Z'}$ and (b) $M_{N_1} = 100 M_{Z'}$.
The $N_1$ becomes thermal in the region above the dashed lines, and non-thermal in the region below the same lines.
The blue curves indicate the points where the DM relic density $\Omega_{N_1} h^2 = 0.12$ is satisfied.}
\label{fig:fig_Oh2}
\end{figure*}

\subsection{$M_{Z'}<2M_{N_1}$ case}
\label{subsec:dm2}
Next, let us further focus on a possible dark matter scenario for $M_{Z'}<2M_{N_1}$, where the BBN bound gets relaxed significantly because the reaction (a) is suppressed.
This can be seen in the region $M_{Z'}\lesssim 20~\kev$ in Fig. \ref{fig:10keV_2}, where the BBN bound on the gauge coupling becomes weak since the $N_1$ is hardly thermalized.
In our setup, there are two scenarios for the dark matter depending on whether the relic abundance is produced in a thermal or non-thermal way.

Figure~\ref{fig:fig_Oh2} shows the $N_1$ relic density for a couple of examples of the $M_{Z'}<2M_{N_1}$ case.
In the region above dashed curves in Fig.~\ref{fig:fig_Oh2}, the $N_1$ comes into the thermal bath at some time.
In this parameter region, we find numerically the $N_1$ is always non-relativistic at the decoupling temperature $T^{\rm dec}_{N_1}$, and thus, we can evaluate the dark matter abundance in the same way as the usual cold dark matter case, which is given by
\begin{eqnarray}
  \Omega_{N_1}^{\rm th}h^2
  &=&
  \frac{s_0 M_{N_1}Y_{N_1}^{\rm th}}{\rho_ch^{-2}},\\
  1/Y_{N_1}^{\rm th}
  &=&
  \left[\frac{45}{8\pi^2 M_{\rm Pl}^2}\right]^{-1/2}\int_0^{T_{N_1}^{\rm dec}}g_*^{1/2}\langle\sigma v\rangle dT,
\end{eqnarray}
where the thermally averaged annihilation cross section, $\langle\sigma v\rangle$, includes the processes (a) and (c).
The gray regions above the dashed curves in Fig.~\ref{fig:fig_Oh2} show the parameter space of $\Omega_{N_1}^{\rm th} h^2 > 0.12$, where we have given two benchmark cases, $M_{N_1}=M_{Z'}$ [Fig.~\ref{fig:fig_Oh2_1}] and $M_{N_1}=100 M_{Z'}$ [Fig.~\ref{fig:fig_Oh2_2}].
In both cases, however, the thermal dark matter scenario is ruled out by various experiments such as the Borexino.\footnote{The thermal $N_1$ dark matter scenario is still viable for $M_{Z'}>2M_{N_1}$ case as mentioned in section \ref{subsec:dm1}.}

As a viable dark matter scenario, let us consider the non-thermal case where the $N_1$ is produced by the freeze-in mechanism discussed earlier.
By demanding the condition $n_{N_1}(T_R)\simeq n_{N_1}(\infty)=0$, we obtain the abundance given by
\begin{eqnarray}
  \Omega_{N_1}^{\rm nt}h^2
  &=&
  \frac{s_0 M_{N_1}Y_{N_1}^{\rm nt}}{\rho_ch^{-2}},\\
  1/Y_{N_1}^{\rm nt}
  &=&
  \left[\frac{45}{8\pi^2 M_{\rm Pl}^2}\right]^{-1/2}\int_0^{\infty}g_*^{1/2}\langle\sigma v\rangle dT. \label{eq:yieldNt}
\end{eqnarray}

An important feature of this case is that the abundance is almost independent from the $N_1$ mass.
To see this, let us approximately derive the analytical expression of the relic abundance.
Since we consider the off-resonance reactions here and only the reaction (a) is sufficient in most of the cases, we can take $\sigma v \sim g_{B-L}^4/(3\pi s)$.
Substituting $\sigma v$ to the right hand side of Eq. (\ref{eq:yieldNt}), we obtain
\begin{eqnarray}
  \frac{dY_{N_1}^{\rm nt}}{dT}
  &=&
  -\frac{45\sqrt{10}}{32\pi^8 g_*^{3/2}}\frac{g_{B-L}^4M_{\rm Pl}M_{N_1}^2}{T^4}K_1^2(M_{N_1}/T).
  \label{eq:Boltzmann3}
\end{eqnarray}
It should be noted that the right hand side of Eq. (\ref{eq:Boltzmann3}) takes the maximum value around $T\sim M_{N_1}$, and thus, the produced number density is not sensitive to higher temperatures.
Because of this, it turns out to be $Y_{N_1}\propto 1/M_{N_1}$ after integrating over the temperature, and hence the abundance is independent of $M_{N_1}$.
By replacing $T$ by $x\equiv M_{N_1}/T$ and integrating over $x$ from $0$ to $\infty$ in Eq. (\ref{eq:Boltzmann3}), we end up with the non-thermal abundance
\begin{eqnarray}
  \Omega_{N_1}^{\rm nt}h^2 
  &\simeq&
  0.12\times\left(\frac{100}{g_*}\right)^{3/2}\left(\frac{g_{B-L}}{4.5\times10^{-6}}\right)^4.
  \label{eq:Oh2nt}
\end{eqnarray}
This estimate well coincides with our numerical calculation shown as the gray regions below the dashed curves in Fig. \ref{fig:fig_Oh2}, where the small fluctuations are caused by the temperature dependence of $g_*$ whose value is roughly given by $g_*(T\sim {\rm max}\{M_{Z'},M_{N_1}\})$.

We briefly comment on the BBN bound in Fig. \ref{fig:fig_Oh2}, which is depicted by the dark blue regions.
Since the BBN bound is sensitive only for the relativistic spices at around the neutrino decoupling temperature, it can eliminate up to $M_{N_1},M_{Z'}\lesssim T^{\rm dec}_{\nu}$.
Below $g_{B-L}\sim 10^{-5}$, the thermalization temperature of the $N_1$ and $Z'$ is below $T^{\rm dec}_{\nu}$ or they never come into thermal bath, and thus the BBN can not constrain this region.

Before closing this section, we note perturbative unitarity on the coupling $\kappa_i$ for $i=2,3$, which can be expressed as $\kappa_i\sim g_{B-L}(M_{N_i}/M_{N_1})(M_{N_1}/M_{Z'})$, and $\kappa_1<\kappa_2,\kappa_3$ should hold as the $N_1$ is the lightest among the three in our setup.
By demanding $\kappa_i<4\pi$, the masses of $N_2$ and $N_3$ are bounded from above as $M_{N_i}/M_{N_1} < (4\pi/g_{B-L})(M_{Z'}/M_{N_1})$.
This is relevant in the case of $M_{N_1}=100 M_{Z'}$ for instance, where we have $M_{N_i}/M_{N_1}\lesssim 0.13/g_{B-L}$.
Namely, when $g_{B-L}$ becomes $g_{B-L}\gsim0.1$, our assumption of taking $M_{N_1}\ll M_{N_2},M_{N_3}$ would be no longer valid.

\begin{figure*}[t]
\begin{center}
\subfigure[]{\label{fig:fig_focus_1}
\includegraphics[width=0.45\textwidth,clip]{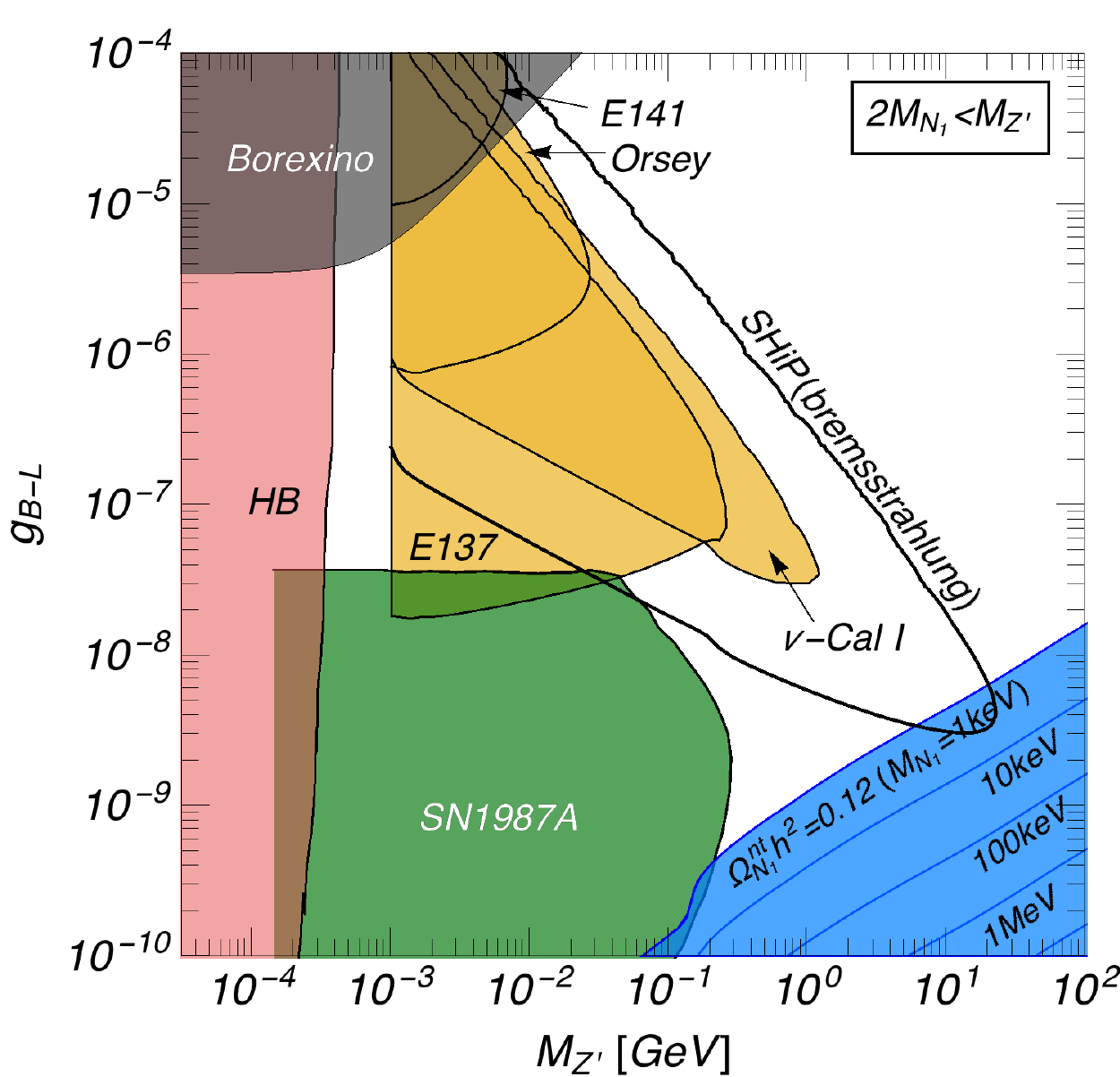}} ~~ 
\subfigure[]{\label{fig:fig_focus_2}
\includegraphics[width=0.45\textwidth,clip]{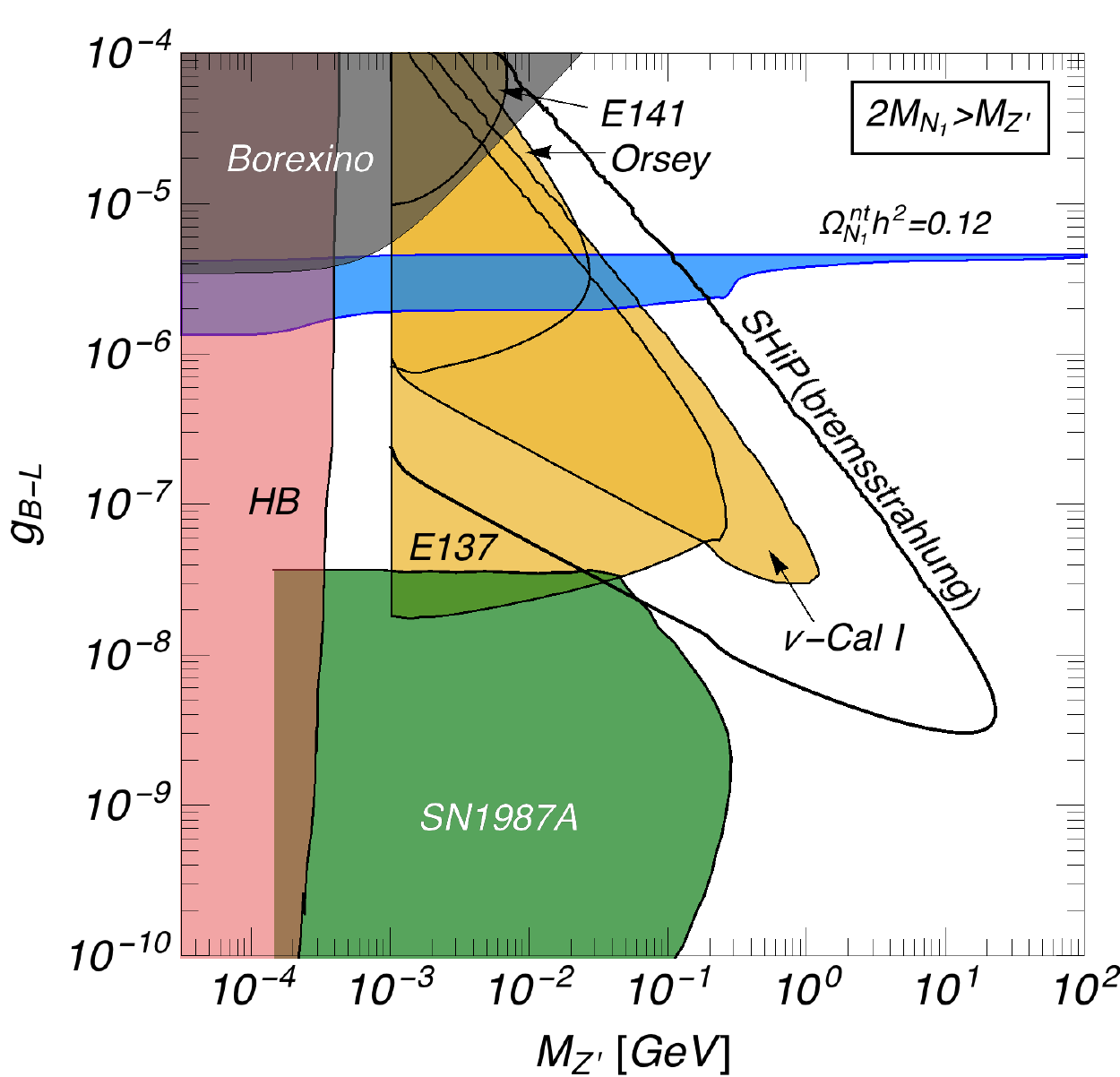}}
\end{center}
\caption{The parameter space around the beam dump constraints and the SHiP sensitivity. The dark matter abundance and other various constraints were imposed.
(a) In the blue area, the correct relic density is achieved for $M_{Z'}>2M_{N_1}$ ($M_{N_1}=10~\kev,~100~\kev,~1~\mev,~10~\mev$ are shown by the blue curves), where we utilized Eq.~(\ref{eq:Oh2nt_res}) taking $C_f=43/6$ and $g_*=g_*(M_{Z'})$.
(b) To show the blue band, which results in the correct relic density, we utilized the estimate given in Eq.~\eqref{eq:Oh2nt}, varying $M_{N_1}$ for all values larger than $M_{Z'} / 2$.
The (a) reflects partially some properties shown in Fig.~\ref{fig:10keV}, and the (b) reflects some properties shown in Fig.~\ref{fig:fig_Oh2}.}
\label{fig:fig_focus}
\end{figure*}

\section{Implications}
\label{sec:implications}
In the non-thermal scenario for $2M_{N_1}>M_{Z'}$, the dark matter abundance given by Eq. (\ref{eq:Oh2nt}) implies the $B-L$ breaking scale.
Since the $Z'$ mass is given by Eq. (\ref{eq:MZp}), substituting Eq. (\ref{eq:Oh2nt}) we end up with the $B-L$ breaking scale $v_S$:
\begin{eqnarray}
  v_S^2
  &\simeq&
  (7.9\times10^4 M_{Z'})^2\left(\frac{0.12}{\Omega_{N_1}^{\rm nt}h^2}\right)^{1/2}\left(\frac{100}{g_*}\right)^{3/4}.
\end{eqnarray}
It turns out that, e.g., for the mass regions $500~\kev\lesssim M_{Z'}\lesssim 1~\mev$ and $M_{Z'}\gtrsim 0.1~\gev$, the scalar mass is at most $200~\gev\lesssim M_S\lesssim 400~\gev$ and $M_S\gtrsim 4~\tev$, respectively, when we take the perturbatively allowed maximum value $\lambda_S=4\pi$.
Although scrutinizing the effect of the $S$ is beyond the scope of this paper, our analysis is valid when we take $\lambda_{HS}$ vanishingly small so that the $S$ does not come into the thermal bath and the non-thermal production of the $N_1$ through the decay of the $S$ is sufficiently small \cite{Asaka:2006ek}.

For direct searches of the dark matter, the scattering between the $N_1$ and a nucleon can be induced by the $Z'$ and $S$ mediated processes.
However, since an effective operator of the $Z'$ mediation is given by $(\overline N_1 \gamma^5\gamma^\mu N_1)(\bar q\gamma_\mu q)$, the scattering cross section is suppressed by velocity or momentum in the non-relativistic limit \cite{Freytsis:2010ne}, which makes the measurement of this process difficult.
We do not consider the $S$ mediated process \cite{Okada:2010wd,Okada:2012sg} as this interaction is turned off by taking $\lambda_{HS}\simeq 0$ in this paper.

Next, let us discuss possible experiments to test the {\it freeze-in} region with the right relic density (roughly, $g_{B-L} \sim 10^{-6}$ region) in Fig. \ref{fig:fig_Oh2}.
Beam dump experiments are a powerful tool to look for the small coupling regions.
We estimate the expected reach of the SHiP experiment \cite{Alekhin:2015byh} using only the proton bremsstrahlung mode.
The SHiP experiment utilizes the CERN SPS 400 GeV proton beam, where the $Z'$ can be produced via bremsstrahlung in proton scattering off the target.
The SHiP is designed to have a 60 m muon shield and a 50 m detector area, and the $Z'$ decaying into the dileptons inside the detector may be observed.
To estimate the signal events, we take the same kinematic parameters shown in Ref.~\cite{Gorbunov:2014wqa}.
Following a similar approach to Ref.~\cite{Gorbunov:2014wqa}, we take no background, in all Figs. \ref{fig:10keV} - \ref{fig:fig_focus} we show the expected region with the signal events more than three, which is depicted by the black solid curves.\footnote{The actual SHiP experiment sensitivity curves will be somewhat different from the ones given in our figures for the higher $Z'$ mass region as they should include additional production channels and the parton level analysis.} \footnote{A study on how the decaying $N_{2,3}$ signals with the $B-L$ gauge boson can appear in the experimental searches at the LHC and the SHiP can be found in Ref.~\cite{Batell:2016zod}.}

Figure~\ref{fig:fig_focus} shows the region around the BD constraints.
In the blue area in Fig.~\ref{fig:fig_focus_1}, the non-thermally produced $N_1$ can explain the correct relic density with the $N_1$ with $M_{N_1}<M_{Z'}/2$.
The SHiP might barely have a chance to test this case when the $M_{N_1}$ is near 10 keV.
The blue band in Fig.~\ref{fig:fig_focus_2} shows the case that the $N_1$ with $M_{N_1}>M_{Z'}/2$ can explain the whole amount of the observed DM abundance.
The bottom side of this band is determined by taking $M_{N_1}\simeq M_{Z'}/2$, and the top side by taking $M_{N_1}\gg M_{Z'}$.
As a result, for the freeze-in region, the SHiP is expected to cover the mass range of $1~\mev\lesssim M_{Z'}\lesssim 200~\mev$.
The two cases in Fig.~\ref{fig:fig_focus_1} and Fig.~\ref{fig:fig_focus_2} are distinguishable in the sense that the $Z'$ decaying into a pair of the light DM is applicable in the former but not in the latter.
While the blue area in the former case is not easily accessible with the planned experiments, the blue band in the latter case is quite well accessible partly because its coupling is larger.

There are some other forthcoming experiments that might be sensitive to our scenario.
The NA64, one of the beam dump experiments at the CERN SPS looks for a missing energy carried away by a light gauge boson \cite{Andreas:2013lya}, and it may be sensitive to the $M_{Z'} < 2 M_e$ region too as the $Z'$ can decay into the neutrinos in the $B-L$ model.
Also the Belle II experiments using the mono-photon trigger has a sensitivity that can cover 10 times smaller than the BABAR results in terms of the gauge coupling \cite{Ferber:2015jzj}.
Detailed analysis for these experiments and developing methods to cover the whole blue regions in Fig.~\ref{fig:fig_focus} are called for.

\section{Summary and Outlook}
\label{sec:summary}
Success of the SM has been astonishing and it is amazingly consistent with high precision experiments.
Yet, there are reasons to believe the complete description of nature requires the SM to be extended.

Following the success of the gauge principle in the SM, we have investigated the minimal gauge $U(1)_{B-L}$ extension of the SM, where three RHNs, a $U(1)_{B-L}$ gauge boson, and a singlet scalar are introduced.
In particular, we have discussed the possibility that the lightest RHN $N_1$ is a dark matter candidate.
Due to the presence of the $Z'$ interaction, the production mechanism of the dark matter does not need to rely on the mixing between active and sterile neutrinos, i.e., Dodelson-Widrow mechanism, and thus the $X$-ray bounds can be evaded.

For the keV scale dark matter, the $U(1)_{B-L}$ gauge interaction can bring the $N_1$ into the thermal bath, and thus the dark matter abundance is determined by the freeze-out mechanism.
The produced $N_1$ is, however, relativistic at its decoupling in most parameter regions, which requires extra entropy production to dilute the overproduced number density.
Note that even if the $N_1$ is never thermalized, non-thermal production such as the freeze-in mechanism can work.
However, the produced number density is fairly small in this case.

As another viable possibility, we have considered heavier mass scales for the $N_1$ DM candidate based on two different relative mass spectrum: $2M_{N_1}<M_{Z'}$ and $2M_{N_1}>M_{Z'}$.

For the $2M_{N_1}<M_{Z'}$ case, we have discussed the freeze-in production of the $N_1$, and found that extremely small $g_{B-L}$ is required for the correct number density for the DM candidate, which makes it difficult to be covered by the planned experiments except for a tiny region in the parameter space.

For the $2M_{N_1}>M_{Z'}$ case, the $N_1$ can be produced either in a thermal or non-thermal way depending on the parameter region.
In the parameter region where the $N_1$ is thermalized, it is always non-relativistic at its decoupling, and thus becomes thermally produced cold dark matter in a typical way.
The thermal abundance, however, requires a rather large gauge coupling, and such regions are already excluded by various experiments (Borexino, etc.).
On the other hand, a non-thermal production is still allowed for a smaller gauge coupling.
We found that the appropriate value of the DM abundance can be obtained for $g_{B-L}\sim 10^{-6}$ largely independent of the $N_1$ mass.
Interestingly, this parameter region (indicated as the blue band in Fig.~\ref{fig:fig_focus_2}) can be sensitive to the planned beam dump experiments, and we found this freeze-in scenario can be tested by the light gauge boson searches at the SHiP experiment up to $M_{Z'}\sim 200~\mev$.

We recall that the muon $g-2$ favored parameter region (of the mass and coupling) in the dark photon scenario \cite{Gninenko:2001hx,Fayet:2007ua,Pospelov:2008zw} has been a target of the active experimental searches in the past decade \cite{Essig:2013lka}.
The parameter space was completely excluded by 2015 through the collaborative efforts of many different experiments \cite{Batley:2015lha}.
The blue band in our study is specifically determined parameter region in our scenario (for the case the $Z'$ does not decay into a pair of the light DM), and some part of it is testable with the planned experiments.
It would be worth investigating the possible ways to completely cover this parameter region of the $B-L$ gauge boson, motivated by the the relic dark matter, the neutrino mass, and the BAU.

We have not scrutinized the interaction through an additional Higgs singlet assuming that $\lambda_{HS}$ is vanishingly small so that it does not contribute to the dark matter production.
We also have not discussed the effect of the $\theta_1$, taking it negligibly small.
The effect of these additional contributions will be discussed elsewhere.

\appendix
\section{Reaction rates}
\label{sec:appendix}
We summarize the formulae used to calculate the relic abundance.
The relevant processes are
\begin{eqnarray}
  {\rm (a)} &~~~& N_1\overline N_1 \leftrightarrow f\bar f,\\
  {\rm (b)} & & Z'Z' \leftrightarrow f\bar f,\\
  {\rm (c)} & & N_1 \overline N_1 \leftrightarrow Z'Z'.
\end{eqnarray}
The squared amplitudes of these processes are given by
\begin{eqnarray}
  \sum_{\rm spins}|{\cal M}_a|^2
  &=&
  \frac{4g_{B-L}^4Q'^2_fN_Cs^2}{(s-M_{Z'}^2)^2+M_{Z'}^2\Gamma_{Z'}^2}
  \left[
  \frac{t^2+u^2}{s^2}-4\frac{M_f^2}{s}\frac{t+u}{s}+2\frac{-M_{N_1}^4-2M_{N_1}^2M_f^2+3M_f^4}{s^2}
  \right], \label{Ma}\\
  \sum_{\rm spins}|{\cal M}_b|^2
  &=&
  \frac{8g_{B-L}^4Q'^4_f tu}{(t-M_f^2)^2}
  \left[
  1-M_f^2\frac{3t+u}{tu}-\frac{M_{Z'}^4+4M_{Z'}^2M_f^2+M_f^4}{tu}
  \right]+ (t\leftrightarrow u)\nn\\
  &&
  -\frac{16g_{B-L}^4Q'^4_f s^2}{(t-M_f^2)(u-M_f^2)}
  \left[
  \frac{2M_{Z'}^2+M_f^2}{s}+
  2\frac{(M_{Z'}^2+2M_f^2)M_f^2}{s^2}
  \right],\label{Mb}\\
    \sum_{\rm spins}|{\cal M}_c|^2
  &=&
  \frac{g_{B-L}^4tu}{(t-M_{N_1}^2)^2}
  \left[
  1-M_{N_1}^2\frac{19t-u}{tu}-\frac{M_{Z'}^4-12M_{Z'}^2M_{N_1}^2+17M_{N_1}^4}{tu}
  \right] + (t\leftrightarrow u)
  \nn\\
  &&
  +\frac{2g_{B-L}^4s^2}{(t-M_{N_1}^2)(u-M_{N_1}^2)}
  \left[\frac{2M_{Z'}^2-3M_{N_1}^2}{s}+2\frac{(6M_{N_1}^2-M_{Z'}^2)M_{N_1}^2}{s^2}\right],
  \label{Mc}
\end{eqnarray}
where $M_f$ represents the SM particle masses, $s,~t,~u$ are the Mandelstam variables, and $N_C$ is the color factor ($N_C=3$ for quarks, otherwise $N_C=1$).
All the squared amplitudes are summed over spins.
For the left-handed neutrinos, we take the massless limit in our numerical analysis. In particular, under this limit, the squared amplitudes of $N_1\overline N_1\leftrightarrow \nu\bar\nu$ and $Z'Z'\leftrightarrow \nu\bar\nu$ become a half of Eq. (\ref{Ma}) and Eq. (\ref{Mb}), respectively.
It should be mentioned that the full expression of the $|{\cal M}_c|^2$ would contain both the longitudinal component contribution which diverges in the high energy region, and the $S$ contribution which cancels the divergence.
Since the dark matter production discussed in this paper is not sensitive to the high energy behavior of ${\cal M}_c$, we did not include them in Eq.~\eqref{Mc}.
They will be presented and discussed in the subsequent work when we discuss the $S$ contribution in detail.

The total decay width of $Z'$ is written by $\Gamma_{Z'}$ of which the hadronic decay channel is obtained by $\Gamma(Z'\to{\rm hadrons})=\Gamma(Z'\to\mu^+\mu^-)R(s=M_{Z'}^2)$ with $R(s)$ being the usual $R$ ratio (at a collision energy $\sqrt{s}$) defined by $R(s)=\sigma(e^+e^-\to{\rm hadrons})/\sigma(e^+e^-\to\mu^+\mu^-)$ \cite{Batell:2009yf}.
The partial decay widths are given by
\begin{eqnarray}
\Gamma(Z'\to f\bar f) &=& \frac{g_{B-L}^2{N_C}Q'^2_fM_{Z'}}{12\pi}\left[1+\frac{2M_f^2}{M_{Z'}^2}\right]\left[1-\frac{4M_f^2}{M_{Z'}^2}\right]^{1/2} , \\
\Gamma(Z'\to N_1\overline N_1) &=& \frac{g_{B-L}^2M_{Z'}}{24\pi}\left[1-\frac{4M_{N_1}^2}{M_{Z'}^2}\right]^{3/2} .
\end{eqnarray}
For the decay of $Z'$ into three massless neutrinos, its partial decay width becomes $\Gamma(Z'\to \nu\bar\nu)=3g_{B-L}^2M_{Z'}/(24\pi)$.

The reaction rates can be defined by using thermally averaged cross sections.
For instance, the reaction rate of the process $N_1\overline N_1\to f\bar f$ is given by $r_a=\langle \sigma v(N_1\overline N_1\to f\bar f)\rangle\times n_{N_1}^{\rm eq}$ where $n_i^{\rm eq}=g_i(2\pi^2)^{-1}M_i^2T K_2(M_i/T)$ is the number density of particle $i$ (having the mass $M_i$ and the degrees of freedom $g_i$, e.g., $g_N=2$ and $g_{Z'}=3$) in the equilibrium state.
($K_2$ is the modified Bessel function of the second kind.)

\begin{acknowledgments}
The work of KK and HL was supported by IBS (Project Code IBS-R018-D1).
We thank J. Heeck for helpful discussions.
HL thanks H. Davoudiasl and W. Marciano for long-term collaboration on the light gauge boson.
We thank conversations with K.C. Kong in the early stage of the project.
\end{acknowledgments}



\begin{thebibliography}{99}

\bibitem{seesaw}
  T. ~Yanagida,
  in Proceedings of the Workshop on Unified Theory and Baryon Number of the Universe,
  eds.  O. Sawada and A. Sugamoto (KEK, 1979) p.95;
  M. Gell- Mann, P. Ramond and R. Slansky,
  in Supergravity,
  eds. P. van Niewwenhuizen and D. Freedman (North Holland, Amsterdam, 1979); 
  S.L. Glashow,
  in Quarks and Leptons, Carg\`{e}se 1979,
  eds. M. L\'{e}vy, et al., (Plenum 1980 New York), p. 707.
  See also 
  P.~Minkowski,
  Phys.\ Lett.\  {\bf B67}, 421 (1977).

\bibitem{Asaka:2005an} 
  T.~Asaka, S.~Blanchet and M.~Shaposhnikov,
  Phys.\ Lett.\ B {\bf 631}, 151 (2005)
  doi:10.1016/j.physletb.2005.09.070
  [hep-ph/0503065].
\bibitem{Asaka:2005pn} 
  T.~Asaka and M.~Shaposhnikov,
  Phys.\ Lett.\ B {\bf 620}, 17 (2005)
  doi:10.1016/j.physletb.2005.06.020
  [hep-ph/0505013].

\bibitem{Boyarsky:2009ix} 
  A.~Boyarsky, O.~Ruchayskiy and M.~Shaposhnikov,
  Ann.\ Rev.\ Nucl.\ Part.\ Sci.\  {\bf 59}, 191 (2009)
  doi:10.1146/annurev.nucl.010909.083654
  [arXiv:0901.0011 [hep-ph]].

\bibitem{Alekhin:2015byh} 
  S.~Alekhin {\it et al.},
  arXiv:1504.04855 [hep-ph].

\bibitem{Adhikari:2016bei} 
  M.~Drewes {\it et al.},
  [arXiv:1602.04816 [hep-ph]].

\bibitem{Akhmedov:1998qx} 
  E.~K.~Akhmedov, V.~A.~Rubakov and A.~Y.~Smirnov,
  Phys.\ Rev.\ Lett.\  {\bf 81}, 1359 (1998)
  doi:10.1103/PhysRevLett.81.1359
  [hep-ph/9803255].

\bibitem{Dodelson:1993je} 
  S.~Dodelson and L.~M.~Widrow,
  Phys.\ Rev.\ Lett.\  {\bf 72}, 17 (1994)
  doi:10.1103/PhysRevLett.72.17
  [hep-ph/9303287].
\bibitem{Barbieri:1989ti} 
  R.~Barbieri and A.~Dolgov,
  Phys.\ Lett.\ B {\bf 237}, 440 (1990).
  doi:10.1016/0370-2693(90)91203-N
\bibitem{Asaka:2006nq} 
  T.~Asaka, M.~Laine and M.~Shaposhnikov,
  JHEP {\bf 0701}, 091 (2007)
  Erratum: [JHEP {\bf 1502}, 028 (2015)]
  doi:10.1088/1126-6708/2007/01/091, 10.1007/JHEP02(2015)028
  [hep-ph/0612182].

\bibitem{Essig:2013goa} 
  R.~Essig, E.~Kuflik, S.~D.~McDermott, T.~Volansky and K.~M.~Zurek,
  JHEP {\bf 1311}, 193 (2013)
  doi:10.1007/JHEP11(2013)193
  [arXiv:1309.4091 [hep-ph]].

\bibitem{Horiuchi:2013noa} 
  S.~Horiuchi, P.~J.~Humphrey, J.~Onorbe, K.~N.~Abazajian, M.~Kaplinghat and S.~Garrison-Kimmel,
  Phys.\ Rev.\ D {\bf 89}, no. 2, 025017 (2014)
  doi:10.1103/PhysRevD.89.025017
  [arXiv:1311.0282 [astro-ph.CO]].

\bibitem{Viel:2013apy} 
  M.~Viel, G.~D.~Becker, J.~S.~Bolton and M.~G.~Haehnelt,
  Phys.\ Rev.\ D {\bf 88}, 043502 (2013)
  doi:10.1103/PhysRevD.88.043502
  [arXiv:1306.2314 [astro-ph.CO]].

\bibitem{Shi:1998km} 
  X.~D.~Shi and G.~M.~Fuller,
  Phys.\ Rev.\ Lett.\  {\bf 82}, 2832 (1999)
  doi:10.1103/PhysRevLett.82.2832
  [astro-ph/9810076].

\bibitem{Asaka:2006ek} 
 M. Shaposhnikov and I. Tkachev, Phys. Lett. B639, 414 (2006); A. Kusenko, Phys. Rev. Lett. 97, 241301 (2006); K. Petraki and A. Kusenko, Phys. Rev. D 77, 065014 (2008);  
  H.~Matsui and M.~Nojiri,
  Phys.\ Rev.\ D {\bf 92}, no. 2, 025045 (2015)
  doi:10.1103/PhysRevD.92.025045
  [arXiv:1503.01293 [hep-ph]].

\bibitem{Merle:2013wta} 
  A.~Merle, V.~Niro and D.~Schmidt,
  JCAP {\bf 1403}, 028 (2014)
  doi:10.1088/1475-7516/2014/03/028
  [arXiv:1306.3996 [hep-ph]];
  Z.~Kang,
  Eur.\ Phys.\ J.\ C {\bf 75}, no. 10, 471 (2015)
  doi:10.1140/epjc/s10052-015-3702-4
  [arXiv:1411.2773 [hep-ph]];
  S.~B.~Roland, B.~Shakya and J.~D.~Wells,
  Phys.\ Rev.\ D {\bf 92}, no. 11, 113009 (2015)
  doi:10.1103/PhysRevD.92.113009
  [arXiv:1412.4791 [hep-ph]];
  A.~Merle and M.~Totzauer,
  JCAP {\bf 1506}, 011 (2015)
  doi:10.1088/1475-7516/2015/06/011
  [arXiv:1502.01011 [hep-ph]];
  Z.~Kang,
  Phys.\ Lett.\ B {\bf 751}, 201 (2015)
  doi:10.1016/j.physletb.2015.10.031
  [arXiv:1505.06554 [hep-ph]];
  A.~Adulpravitchai and M.~A.~Schmidt,
  JHEP {\bf 1512}, 023 (2015)
  doi:10.1007/JHEP12(2015)023
  [arXiv:1507.05694 [hep-ph]];
  M.~Drewes and J.~U.~Kang,
  JHEP {\bf 1605}, 051 (2016)
  doi:10.1007/JHEP05(2016)051
  [arXiv:1510.05646 [hep-ph]].

\bibitem{FIMP} 
  J.~McDonald,
  Phys.\ Rev.\ Lett.\  {\bf 88}, 091304 (2002)
  doi:10.1103/PhysRevLett.88.091304
  [hep-ph/0106249];
  L.~J.~Hall, K.~Jedamzik, J.~March-Russell and S.~M.~West,
  JHEP {\bf 1003}, 080 (2010)
  doi:10.1007/JHEP03(2010)080
  [arXiv:0911.1120 [hep-ph]].

\bibitem{Chu:2011be} 
  X.~Chu, T.~Hambye and M.~H.~G.~Tytgat,
  JCAP {\bf 1205}, 034 (2012)
  doi:10.1088/1475-7516/2012/05/034
  [arXiv:1112.0493 [hep-ph]];
  X.~Chu, Y.~Mambrini, J.~Quevillon and B.~Zaldivar,
  JCAP {\bf 1401}, 034 (2014)
  doi:10.1088/1475-7516/2014/01/034
  [arXiv:1306.4677 [hep-ph]].


\bibitem{ADM}
  E.~J.~Chun,
  JHEP {\bf 1103}, 098 (2011)
  doi:10.1007/JHEP03(2011)098
  [arXiv:1102.3455 [hep-ph]];
  M.~Ibe, S.~Matsumoto and T.~T.~Yanagida,
  Phys.\ Lett.\ B {\bf 708}, 112 (2012)
  doi:10.1016/j.physletb.2012.01.032
  [arXiv:1110.5452 [hep-ph]];
  K.~Petraki, M.~Trodden and R.~R.~Volkas,
  JCAP {\bf 1202}, 044 (2012)
  doi:10.1088/1475-7516/2012/02/044
  [arXiv:1111.4786 [hep-ph]];
  N.~Okada and O.~Seto,
  Phys.\ Rev.\ D {\bf 86}, 063525 (2012)
  doi:10.1103/PhysRevD.86.063525
  [arXiv:1205.2844 [hep-ph]];
  W.~Z.~Feng and P.~Nath,
  Phys.\ Lett.\ B {\bf 731}, 43 (2014)
  doi:10.1016/j.physletb.2014.02.020
  [arXiv:1312.1334 [hep-ph]].

\bibitem{Okada:2016gsh} 
  N.~Okada and S.~Okada,
  Phys.\ Rev.\ D {\bf 93}, no. 7, 075003 (2016)
  doi:10.1103/PhysRevD.93.075003
  [arXiv:1601.07526 [hep-ph]].

\bibitem{Okada:2010wd} 
  N.~Okada and O.~Seto,
  Phys.\ Rev.\ D {\bf 82}, 023507 (2010)
  doi:10.1103/PhysRevD.82.023507
  [arXiv:1002.2525 [hep-ph]].
\bibitem{Okada:2012sg} 
  N.~Okada and Y.~Orikasa,
  Phys.\ Rev.\ D {\bf 85}, 115006 (2012)
  doi:10.1103/PhysRevD.85.115006
  [arXiv:1202.1405 [hep-ph]].

\bibitem{Guo:2015lxa} 
  J.~Guo, Z.~Kang, P.~Ko and Y.~Orikasa,
  Phys.\ Rev.\ D {\bf 91}, no. 11, 115017 (2015)
  doi:10.1103/PhysRevD.91.115017
  [arXiv:1502.00508 [hep-ph]].

\bibitem{Bezrukov:2009th} 
  F.~Bezrukov, H.~Hettmansperger and M.~Lindner,
  Phys.\ Rev.\ D {\bf 81}, 085032 (2010)
  doi:10.1103/PhysRevD.81.085032
  [arXiv:0912.4415 [hep-ph]].
  
\bibitem{Nemevsek:2012cd} 
  M.~Nemevsek, G.~Senjanovic and Y.~Zhang,
  JCAP {\bf 1207}, 006 (2012)
  doi:10.1088/1475-7516/2012/07/006
  [arXiv:1205.0844 [hep-ph]].

\bibitem{Holdom:1985ag} 
  B.~Holdom,
  Phys.\ Lett.\ B {\bf 166}, 196 (1986).
  doi:10.1016/0370-2693(86)91377-8

\bibitem{Essig:2013lka} 
  R.~Essig {\it et al.},
  arXiv:1311.0029 [hep-ph].

\bibitem{Davoudiasl:2012ag} 
  H.~Davoudiasl, H.~S.~Lee and W.~J.~Marciano,
  Phys.\ Rev.\ D {\bf 85}, 115019 (2012)
  doi:10.1103/PhysRevD.85.115019
  [arXiv:1203.2947 [hep-ph]].
     
\bibitem{Lee:2016ief} 
  H.~S.~Lee and S.~Yun,
  Phys.\ Rev.\ D {\bf 93}, no. 11, 115028 (2016)
  doi:10.1103/PhysRevD.93.115028
  [arXiv:1604.01213 [hep-ph]].

\bibitem{Pal:1981rm} 
  P.~B.~Pal and L.~Wolfenstein,
  Phys.\ Rev.\ D {\bf 25}, 766 (1982).
  doi:10.1103/PhysRevD.25.766

\bibitem{Barger:1995ty} 
  V.~D.~Barger, R.~J.~N.~Phillips and S.~Sarkar,
  Phys.\ Lett.\ B {\bf 352}, 365 (1995)
  Erratum: [Phys.\ Lett.\ B {\bf 356}, 617 (1995)]
  doi:10.1016/0370-2693(95)00486-5, 10.1016/0370-2693(95)00831-5
  [hep-ph/9503295].

\bibitem{Agashe:2014kda} 
  K.~A.~Olive {\it et al.} [Particle Data Group Collaboration],
  Chin.\ Phys.\ C {\bf 38}, 090001 (2014).
  doi:10.1088/1674-1137/38/9/090001

\bibitem{Hindmarsh:2005ix} 
  M.~Hindmarsh and O.~Philipsen,
  Phys.\ Rev.\ D {\bf 71}, 087302 (2005)
  doi:10.1103/PhysRevD.71.087302
  [hep-ph/0501232];
  F.~Karsch, E.~Laermann and A.~Peikert,
  Phys.\ Lett.\ B {\bf 478}, 447 (2000)
  doi:10.1016/S0370-2693(00)00292-6
  [hep-lat/0002003].

\bibitem{Ade:2015xua} 
  P.~A.~R.~Ade {\it et al.} [Planck Collaboration],
  arXiv:1502.01589 [astro-ph.CO].

\bibitem{Ahlgren:2013wba} 
  B.~Ahlgren, T.~Ohlsson and S.~Zhou,
  Phys.\ Rev.\ Lett.\  {\bf 111}, no. 19, 199001 (2013)
  doi:10.1103/PhysRevLett.111.199001
  [arXiv:1309.0991 [hep-ph]].

\bibitem{Heeck:2014zfa} 
  J.~Heeck,
  Phys.\ Lett.\ B {\bf 739}, 256 (2014)
  doi:10.1016/j.physletb.2014.10.067
  [arXiv:1408.6845 [hep-ph]].
  
\bibitem{Jeong:2015bbi} 
  Y.~S.~Jeong, C.~S.~Kim and H.~S.~Lee,
  Int.\ J.\ Mod.\ Phys.\ A {\bf 31}, no. 11, 1650059 (2016)
  doi:10.1142/S0217751X16500597
  [arXiv:1512.03179 [hep-ph]].

\bibitem{Carena:2004xs} 
  M.~Carena, A.~Daleo, B.~A.~Dobrescu and T.~M.~P.~Tait,
  Phys.\ Rev.\ D {\bf 70}, 093009 (2004)
  doi:10.1103/PhysRevD.70.093009
  [hep-ph/0408098].

\bibitem{Achard:2003tx} 
  P.~Achard {\it et al.} [L3 Collaboration],
  Phys.\ Lett.\ B {\bf 587}, 16 (2004)
  doi:10.1016/j.physletb.2004.01.010
  [hep-ex/0402002].

\bibitem{Lees:2014xha} 
  J.~P.~Lees {\it et al.} [BaBar Collaboration],
  Phys.\ Rev.\ Lett.\  {\bf 113}, no. 20, 201801 (2014)
  doi:10.1103/PhysRevLett.113.201801
  [arXiv:1406.2980 [hep-ex]].

\bibitem{Bross:1989mp} 
  A.~Bross, M.~Crisler, S.~H.~Pordes, J.~Volk, S.~Errede and J.~Wrbanek,
  Phys.\ Rev.\ Lett.\  {\bf 67}, 2942 (1991).
  doi:10.1103/PhysRevLett.67.2942

\bibitem{Riordan:1987aw} 
  E.~M.~Riordan {\it et al.},
  Phys.\ Rev.\ Lett.\  {\bf 59}, 755 (1987).
  doi:10.1103/PhysRevLett.59.755

\bibitem{Davier:1989wz} 
  M.~Davier and H.~Nguyen Ngoc,
  Phys.\ Lett.\ B {\bf 229}, 150 (1989).
  doi:10.1016/0370-2693(89)90174-3

\bibitem{Blumlein:2013cua} 
  J.~Blümlein and J.~Brunner,
  Phys.\ Lett.\ B {\bf 731}, 320 (2014)
  doi:10.1016/j.physletb.2014.02.029
  [arXiv:1311.3870 [hep-ph]].

\bibitem{Bjorken:1988as} 
  J.~D.~Bjorken {\it et al.},
  Phys.\ Rev.\ D {\bf 38}, 3375 (1988).
  doi:10.1103/PhysRevD.38.3375

\bibitem{Gorbunov:2014wqa} 
  D.~Gorbunov, A.~Makarov and I.~Timiryasov,
  Phys.\ Rev.\ D {\bf 91}, no. 3, 035027 (2015)
  doi:10.1103/PhysRevD.91.035027
  [arXiv:1411.4007 [hep-ph]].

\bibitem{Andreas:2012mt} 
  S.~Andreas, C.~Niebuhr and A.~Ringwald,
  Phys.\ Rev.\ D {\bf 86}, 095019 (2012)
  doi:10.1103/PhysRevD.86.095019
  [arXiv:1209.6083 [hep-ph]].

\bibitem{Bellini:2011rx} 
  G.~Bellini {\it et al.},
  Phys.\ Rev.\ Lett.\  {\bf 107}, 141302 (2011)
  doi:10.1103/PhysRevLett.107.141302
  [arXiv:1104.1816 [hep-ex]].

\bibitem{Harnik:2012ni} 
  R.~Harnik, J.~Kopp and P.~A.~N.~Machado,
  JCAP {\bf 1207}, 026 (2012)
  doi:10.1088/1475-7516/2012/07/026
  [arXiv:1202.6073 [hep-ph]].
  
\bibitem{Bilmis:2015lja} 
  S.~Bilmis, I.~Turan, T.~M.~Aliev, M.~Deniz, L.~Singh and H.~T.~Wong,
  Phys.\ Rev.\ D {\bf 92}, no. 3, 033009 (2015)
  doi:10.1103/PhysRevD.92.033009
  [arXiv:1502.07763 [hep-ph]].
  
\bibitem{Zeller:2001hh} 
  G.~P.~Zeller {\it et al.} [NuTeV Collaboration],
  Phys.\ Rev.\ Lett.\  {\bf 88}, 091802 (2002)
  Erratum: [Phys.\ Rev.\ Lett.\  {\bf 90}, 239902 (2003)]
  doi:10.1103/PhysRevLett.88.091802
  [hep-ex/0110059].
\bibitem{Escrihuela:2011cf} 
  F.~J.~Escrihuela, M.~Tortola, J.~W.~F.~Valle and O.~G.~Miranda,
  Phys.\ Rev.\ D {\bf 83}, 093002 (2011)
  doi:10.1103/PhysRevD.83.093002
  [arXiv:1103.1366 [hep-ph]].

\bibitem{Raffelt:2000kp} 
  G.~G.~Raffelt,
  Phys.\ Rept.\  {\bf 333}, 593 (2000).
  doi:10.1016/S0370-1573(00)00039-9

\bibitem{Redondo:2013lna} 
  J.~Redondo and G.~Raffelt,
  JCAP {\bf 1308}, 034 (2013)
  doi:10.1088/1475-7516/2013/08/034
  [arXiv:1305.2920 [hep-ph]].

\bibitem{Dent:2012mx} 
  J.~B.~Dent, F.~Ferrer and L.~M.~Krauss,
  arXiv:1201.2683 [astro-ph.CO].

\bibitem{Kazanas:2014mca} 
  D.~Kazanas, R.~N.~Mohapatra, S.~Nussinov, V.~L.~Teplitz and Y.~Zhang,
  Nucl.\ Phys.\ B {\bf 890}, 17 (2014)
  doi:10.1016/j.nuclphysb.2014.11.009
  [arXiv:1410.0221 [hep-ph]].


\bibitem{Nelson:2007yq} 
  A.~E.~Nelson and J.~Walsh,
  Phys.\ Rev.\ D {\bf 77}, 033001 (2008)
  doi:10.1103/PhysRevD.77.033001
  [arXiv:0711.1363 [hep-ph]].

\bibitem{Iso:2009ss} 
  S.~Iso, N.~Okada and Y.~Orikasa,
  Phys.\ Lett.\ B {\bf 676}, 81 (2009)
  doi:10.1016/j.physletb.2009.04.046
  [arXiv:0902.4050 [hep-ph]];
  S.~Iso, N.~Okada and Y.~Orikasa,
  Phys.\ Rev.\ D {\bf 80}, 115007 (2009)
  doi:10.1103/PhysRevD.80.115007
  [arXiv:0909.0128 [hep-ph]];

\bibitem{Basso:2008iv} 
  L.~Basso, A.~Belyaev, S.~Moretti and C.~H.~Shepherd-Themistocleous,
  Phys.\ Rev.\ D {\bf 80}, 055030 (2009)
  doi:10.1103/PhysRevD.80.055030
  [arXiv:0812.4313 [hep-ph]].

\bibitem{Gelmini:2004ah} 
  G.~Gelmini, S.~Palomares-Ruiz and S.~Pascoli,
  Phys.\ Rev.\ Lett.\  {\bf 93}, 081302 (2004)
  doi:10.1103/PhysRevLett.93.081302
  [astro-ph/0403323];
  C.~E.~Yaguna,
  JHEP {\bf 0706}, 002 (2007)
  doi:10.1088/1126-6708/2007/06/002
  [arXiv:0706.0178 [hep-ph]];
  G.~Gelmini, E.~Osoba, S.~Palomares-Ruiz and S.~Pascoli,
  JCAP {\bf 0810}, 029 (2008)
  doi:10.1088/1475-7516/2008/10/029
  [arXiv:0803.2735 [astro-ph]];
  S.~Khalil and O.~Seto,
  JCAP {\bf 0810}, 024 (2008)
  doi:10.1088/1475-7516/2008/10/024
  [arXiv:0804.0336 [hep-ph]].


\bibitem{Freytsis:2010ne} 
  M.~Freytsis and Z.~Ligeti,
  Phys.\ Rev.\ D {\bf 83}, 115009 (2011)
  doi:10.1103/PhysRevD.83.115009
  [arXiv:1012.5317 [hep-ph]].

\bibitem{Batell:2016zod} 
  B.~Batell, M.~Pospelov and B.~Shuve,
  arXiv:1604.06099 [hep-ph].

\bibitem{Andreas:2013lya} 
  S.~Andreas {\it et al.},
  arXiv:1312.3309 [hep-ex].

\bibitem{Ferber:2015jzj} 
  T.~Ferber,
  Acta Phys.\ Polon.\ B {\bf 46}, no. 11, 2285 (2015).
  doi:10.5506/APhysPolB.46.2285
  
\bibitem{Gninenko:2001hx} 
  S.~N.~Gninenko and N.~V.~Krasnikov,
  Phys.\ Lett.\ B {\bf 513}, 119 (2001)
  doi:10.1016/S0370-2693(01)00693-1
  [hep-ph/0102222].

\bibitem{Fayet:2007ua} 
  P.~Fayet,
  Phys.\ Rev.\ D {\bf 75}, 115017 (2007)
  doi:10.1103/PhysRevD.75.115017
  [hep-ph/0702176 [HEP-PH]].

\bibitem{Pospelov:2008zw} 
  M.~Pospelov,
  Phys.\ Rev.\ D {\bf 80}, 095002 (2009)
  doi:10.1103/PhysRevD.80.095002
  [arXiv:0811.1030 [hep-ph]].

\bibitem{Batley:2015lha} 
  J.~R.~Batley {\it et al.} [NA48/2 Collaboration],
  Phys.\ Lett.\ B {\bf 746}, 178 (2015)
  doi:10.1016/j.physletb.2015.04.068
  [arXiv:1504.00607 [hep-ex]].
  
\bibitem{Batell:2009yf} 
  B.~Batell, M.~Pospelov and A.~Ritz,
  Phys.\ Rev.\ D {\bf 79}, 115008 (2009)
  doi:10.1103/PhysRevD.79.115008
  [arXiv:0903.0363 [hep-ph]];
  V.~V.~Ezhela, S.~B.~Lugovsky and O.~V.~Zenin,
  hep-ph/0312114.
                            
\end{thebibliography}
\end{document}